\documentclass[%
aip,
 amsmath,amssymb,
 reprint,%
]{revtex4-1}

\usepackage{graphicx}% Include figure files
\usepackage{dcolumn}% Align table columns on decimal point
\usepackage{bm}% bold math
\usepackage[utf8]{inputenc}
\usepackage[T1]{fontenc}
\usepackage{mathptmx}
\usepackage{amsmath}
\usepackage{amssymb}
\usepackage{etoolbox}
\usepackage{ulem}
\usepackage{xcolor}

\makeatletter
\def\@email#1#2{%
 \endgroup
 \patchcmd{\titleblock@produce}
  {\frontmatter@RRAPformat}
  {\frontmatter@RRAPformat{\produce@RRAP{*#1\href{mailto:#2}{#2}}}\frontmatter@RRAPformat}
  {}{}
}%
\makeatother

\begin{document}

\preprint{AIP/123-QED}

\title[Autoencoders for dimensionality reduction in molecular dynamics]{Autoencoders for dimensionality reduction in molecular dynamics:\\ collective variable dimension, biasing and transition states}
% Force line breaks with \\

\author{Zineb Belkacemi$\dag$}
\altaffiliation{Present address: Chipiron, 6 Rue Jean Calvin, Paris, France}
\affiliation{Integrated Drug Discovery, Molecular Design Sciences, Sanofi, Vitry-sur-Seine, France}
\author{Marc Bianciotto$\dag$}
\affiliation{Integrated Drug Discovery, Molecular Design Sciences, Sanofi, Vitry-sur-Seine, France}
\author{Herv\'{e} Minoux}
\altaffiliation{Present address: Digital and Data Sciences, Sanofi, Chilly-Mazarin, France}
\affiliation{Integrated Drug Discovery, Molecular Design Sciences, Sanofi, Vitry-sur-Seine, France}
\author{Tony Leli\`{e}vre}%
\affiliation{CERMICS, Ecole des Ponts, Marne-la-Vallée, France} 
\affiliation{MATHERIALS team-project, Inria Paris, France}
\author{Gabriel Stoltz}%
\affiliation{CERMICS, Ecole des Ponts, Marne-la-Vallée, France} 
\affiliation{MATHERIALS team-project, Inria Paris, France}
\author{Paraskevi Gkeka}
\email{Paraskevi.Gkeka@sanofi.com}
\affiliation{Integrated Drug Discovery, Molecular Design Sciences, Sanofi, Vitry-sur-Seine, France}

\date{\today}% It is always \today, today,

\begin{abstract}
The heat shock protein 90 (Hsp90) is a molecular chaperone that controls the folding and activation of client proteins using the free energy of ATP hydrolysis. The Hsp90 active site is in its N-terminal domain (NTD). Our goal is to characterize the dynamics of NTD using an autoencoder-learned collective variable (CV) in conjunction with adaptive biasing force (ABF) Langevin dynamics. Using dihedral analysis, we cluster all available experimental Hsp90 NTD structures into distinct native states. We then perform unbiased molecular dynamics (MD) simulations to construct a dataset that represents each state and use this dataset to train an autoencoder. Two autoencoder architectures are considered, with one and two hidden layers respectively, and bottlenecks of dimension $k$ ranging from 1 to 10. We demonstrate that the addition of an extra hidden layer does not significantly improve the performance, while it leads to complicated CVs that increases the computational cost of biased MD calculations. In addition, a 2D bottleneck can provide enough information of the different states, while the optimal bottleneck dimension is five. For the 2D bottleneck, the two-dimensional CV is directly used in biased MD simulations. For the 5D bottleneck, we perform an analysis of the latent CV space and identify the pair of CV coordinates that best separates the states of Hsp90. Interestingly, selecting a 2D CV out of the 5D CV space leads to better results than directly learning a 2D CV, and allows to observe transitions between native states when running free energy biased dynamics.
\end{abstract}

\maketitle

\section{\label{sec:intro} Free energy biasing, collective variables and autoencoders}

Efficiently sampling the full configuration space of complex systems in molecular dynamics remains a challenge, due to the long timescales involved in the transition events from one metastable state to another one. Large proteins are one of the historic, and still particularly relevant example of metastable system. Direct numerical simulation of proteins is limited to physical times of a nanoseconds, exceptionally (fraction of) milliseconds.\cite{Lindorff-Larsen2011,Jung2019} Biologically relevant conformational changes are rare events on this timescale. The dynamics needs to be biased in order to favor otherwise unlikely transitions. Metastability is caused by the usually very irregular shape of the energy function that contains many local minima (metastable states) separated by high energy barriers (transition states).

We recall in this section one convenient way to this bias the dynamics and increase the frequency of transition events from one metastable state to another, namely the used of a biasing force corresponding to (an approximation of) the derivative of the free energy. We start by recalling in Section~\ref{sec:FE_biasing} the principle of free energy biasing, and then discuss in Section~\ref{sec:CVs} the key element in this method, namely the choice of the collective variable (CV). This function was chosen mostly based on domain knowledge and physical intuition until a few years ago, but, as we discuss in Section~\ref{sec:AE-CV}, the recent advances in machine learning techniques, and in particular deep learning methods, have revolutionized the way CVs are constructed.

\subsection{Free energy biasing}
\label{sec:FE_biasing}

Adaptive biasing methods are important sampling techniques where the free energy~$F$ is simultaneously estimated and used to bias the potential. More precisely, the biasing relies on the use of a collective variable~$\xi$, which maps the system from the high dimensional molecular space to a much smaller dimensional space~$\mathbb{R}^d$, which effectively describes the metastability of the dynamics. The potential energy function of the system is then replaced by the modified potential $V-F\circ \xi$, where~$F$ is the free energy associated with~$\xi$, in order to eliminate the metastability along~$\xi$ and, thus, help accelerating the sampling of transitions between metastable states. Here and in the following, $\circ$ is the function composition operator i.e.~$F\circ \xi(x) = F(\xi(x))$. We refer to Refs.~\onlinecite{Chipot2007,Lelievre2010,Chipot2014,Henin_Lelievre_Shirts_Valsson_Delemotte_2022,Chipot2023} for instance for reviews and pedagogical introductions to free-energy calculation methods.

However, the free energy~$F$ is of course unknown in general. Adaptive free energy biasing methods replace the free energy~$F$ by an estimated function~$F_t$ in the biased dynamics at time~$t$. The potential becomes~$V-F_t\circ \xi$, where the estimate $F_t$ is updated on-the-fly, and converges to $F$ as the sampling proceeds. There are two categories of adaptive biasing techniques:\cite{LRS2007} (i) Adaptive Biasing Potentials, e.g., Metadynamics,\cite{MetaDyn} where the free energy~$F_t$ is estimated and its gradient, the so-called mean force, is then derived and used in the dynamics; and (ii) Adaptive Biasing Force (ABF) methods,\cite{Darve2001,HC04} where the free energy derivative, i.e., the mean force, is estimated directly as a vector $\Gamma_t$, and the free energy is subsequently obtained by numerical integration of the mean force.\cite{helmholtz}  

By design, ABF requires the knowledge of second order derivatives of the CV~$\xi$ to compute the local mean force $f$. The analytical expression of this quantity is cumbersome for most choices of reaction coordinates, especially when $\xi$ is vector valued. To overcome this limitation a method coined extended system ABF (eABF)\cite{tony} was devised. A fictitious degree of freedom $\lambda$ is added to the configurational space, as in early versions of metadynamics. The new extended mean force does not depend on the second (or any) derivatives of the CV~$\xi$. Only the gradient of~$\xi$ is needed for computing the gradient of~$V^{\text{ext}}$. ABF can therefore easily be applied to the new extended system.

%------------------------------------------------------------------------%
\subsection{\label{sec:CVs} Choosing collective variables}

The collective variable $\xi$ greatly impacts the physical relevance of the computed free energy differences. As mentioned in the introductory lines of this section, the dynamics are often metastable in biomolecular systems. As a consequence, a CV $\xi$ should be chosen to explore these metastable states that often describe key biologically-relevant biomolecular conformations and dynamics. Classical examples of reaction coordinates are combinations of well defined simple functions of the positions~$q$, such as distances, dihedrals or contacts.

Besides, the choice of the CV also impacts the efficiency of the free energy adaptive biasing procedure. When the CV $\xi$ is able to describe the slow motions of interest, the process $(\xi(q_t))$ is also metastable, i.e., its value may stay trapped inside some region of the space $\mathbb{R}^d$ before crossing to another region, indicating a transition of the system from one metastable state to another. As discussed in the previous section, the free energy associated with~$\xi$ can be used to bias the potential of the system so as to make the process $(\xi(q_t))$ no longer metastable. In fact, the marginal distribution along~$\xi$ under the potential $V-F\circ \xi$ is uniform. This motivates even more the importance of the choice of the CV: The biased potential $V-F\circ \xi$ is only as effective at sampling metastable motions of interest as the CV~$\xi$ is at describing them. 

With a poor choice of CV, the free energy cannot provide an efficient biasing of the dynamics and cannot be used for analysing important motions of the system. It is thus primordial to use a CV that encompasses the metastability of the system, a notion which can be mathematically quantified.\cite{lelievre2013two} In general, the choice of the CV can be made somewhat intuitively for small and/or extensively studied systems. However, the more complex and/or larger the system is, the less trivial it is to manually select a CV. The idea of automatically selecting or constructing the collective variable thus becomes attractive. For this purpose, many methods have been devised to construct CVs using sampled configurations of a given system. In particular, as the recent years have known a surge in popularity for machine learning (ML) techniques in various fields, the discovery of collective variables using machine learning has gained growing interest. There are now various reviews on this lively topic, see in particular Refs.~\onlinecite{Ferguson2018,sidkyreview,Gkeka2020,glielmoreview,Chen2021}.

There are two main classes of ML methods to find CVs: those seeking high variances CVs, which aim at reproducing overall features of the Boltzmann--Gibbs distribution at hand; and those seeking slowly evolving CVs (such as tICA\cite{MS94,NF13} for instance). Both classes can be separated into linear and nonlinear methods. Our focus in this work is on high variance CVs. ML-guided methods in this context can be separated into (i) linear algorithms, e.g. PCA or factor analysis, and (ii) non-linear algorithms, e.g. kernel methods, autoencoders, decision trees and random forests (see for instance Refs.~\onlinecite{Mehta2019,Murphy} for introductory references on these classes of methods). In the first case, CVs are interpretable, but often lack of the necessary complexity to describe complex biological phenomena. In the present study, we employ one method of the second type, namely an autoencoder, following up on our previous work.\cite{belkacemi2021}

%------------------------------------------------------------------------%
\subsection{\label{sec:Autoencoders} CV identification using autoencoders}

Artificial Neural Networks (NNs) mimic by design the function of the human brain, in that artificial neurons are made to send signals to one another. More precisely, a NN is composed of several layers, each of which contains a number of neurons. The input layer contains as many neurons as the dimension of the input data and the output layer neurons are meant to contain the information we wish to learn using the NN. The intermediate layers are used to increase the complexity and expressivity of the NN and aid in the learning process. A neuron in each layer is assigned a weight vector which connects it to the neurons of the next layer. The value in each neuron is then computed as a weighted combination of the values of the previous layer neurons, passed through a nonlinear differentiable transformation called activation function. The neural network is optimized by modifying the weights assigned to all neurons, so as to optimize a target distance between the NN's predicted output (i.e. output layer) and the actual output. 

Autoencoders (AE)\cite{AANN} are a type of neural network designed for unsupervised learning tasks.  The aim is usually to learn a new representation of the data, called an encoding. The AE is composed of two parts: the encoder learns the new representation and the decoder simultaneously learns to reconstruct the original data from this representation. The AE thus seeks to approximate the identity function. When the encoder is composed of one fully connected layer which reduces the dimension, together with a linear activation function, its learned representation is essentially the same as that of a PCA projection of the same dimensionality~\cite{encispca1, encispca2}; more precisely, the two models project on the same bottleneck space, but not using the same vectors. In general, however, AEs are used with nonlinear activation functions. This allows for nonlinear encoding functions, and thus potentially better encoders than those restricted to stay within the smaller class of linear functions. 

AEs can have different topologies depending on the learning task, the data size and dimensionality, etc. Below, we describe the general autoencoder topology used in this work. We denote by $\mathcal{X} \subseteq \mathbb{R}^D$ the data space, and by $\mathcal{A} \subseteq \mathbb{R}^d$ a lower dimensional space ($d < D$). The autoencoder can be represented by a mapping $f = f_{\text{dec}} \circ f_{\text{enc}}$, where $f_{\text{enc}}: \mathcal{X} \xrightarrow{} \mathcal{A}$, $f_{\text{dec}}: \mathcal{A} \xrightarrow{} \mathcal{X}$ and $\circ$ is again the function composition operator: $f_{\text{dec}} \circ f_{\text{enc}}(x) = f_{\text{dec}}\left(f_{\text{enc}}(x)\right)$.  The AEs we consider are symmetric in structure, fully connected, and contain~$2L$ layers. Each hidden layer is of dimension $d_{\ell}=d_{2L-\ell}$ for $\ell=1, \dots, L$, and the output layer is of dimension $d_{2L} = D$ (by convention, the input layer does not count as a layer of the network). Each layer $\ell \in {1,\dots,2L}$ has an activation function $g_{\ell}$ and is connected to the previous layer by a projection matrix $W_{\ell} \in \mathbb{R}^{d_{\ell}\times d_{\ell-1}}$, and a bias vector $b_{\ell} \in \mathbb{R}^{d_{\ell}}$. There are thus $K= \displaystyle \sum_{\ell=1}^L d_{\ell} (d_{\ell-1}+1)$ learnable real parameters denoted by $(p_1, \dots, p_K) \in \mathbb{R}^K$. As the activation functions are predefined and do not change during learning, the autoencoder function is fully described by its parameters~$(p_k)_{k=1,\dots,K}$. 

%------------------------------------------------------------------------%
\section{\label{sec:HSP90-structural} Hsp90: an important pharmacological target with many faces}

The heat shock protein 90 (Hsp90) is an ATP-dependent molecular chaperone that controls protein maturation, stability, and folding of over 100 key cellular growth-regulatory and signaling molecules.\cite{Taipale2010} These molecules, i.e., Hsp90 clients, include ~60\% of the human kinome, transcription factors, and multiple mutated, chimeric, and overexpressed signaling proteins that promote cancer cell growth and survival.\cite{Whitesell2005,Zhao2005,Mader2020} These wide variety of biologically critical clients in combination with its interaction with several co-chaperones makes Hsp90 a promising pharmacological target against diseases such as cancer, Alzheimer and other neurodegenerative diseases, diabetes, as well as viral and bacterial infections.\cite{Zuehlke2018,Rowles2020,Zhang2022} 

Hsp90 is a highly conserved enzyme that dimerizes and comprises an N-terminal ATP binding domain (NTD), a middle co-chaperone and client-binding domain (M-domain), and a C-terminal dimerization domain (CTD).\cite{Meyer2003} Interestingly, the Hsp90-NTD possesses an unconventional ATP binding site with a structure named Bergerat fold.\cite{Grenert1997,Dutta2000} This ATP binding pocket is situated within the $\alpha$-helices formed by residues 28-51, 85-97 and 123-130 of NTD of human Hsp90. The mechanism of action of Hsp90 is driven by ATP-influenced large-scale conformational changes during its chaperone cycle.\cite{Wegele2003} These conformational changes are in fact transitions, or even more precisely transition paths, between an open inactive conformation when in apo form, i.e., in the absence of bound nucleotide, to a close active conformation. This exchange between open and closed conformations takes place at a time scale ranging from milliseconds to several minutes.\cite{Berezhkovskii2020} 

Despite the large and slow conformational changes between open and close states during its chaperon activity, the individual Hsp90 domains remain largely stable. Most of the flexibility of Hsp90 is the result of rigid body movements centered at the linkers between the NTD and M-domain, and between the MD and CTD. Interestingly, for these large scale global movements to occur, significant local changes take place in the NTD during nucleotide binding and unbinding.\cite{Krukenberg2011} 

The most significant of these local changes occur in the so-called ``ATP-lid" or ``active site lid", a helix-loop-helix segment adjacent to the ATP-binding pocket.\cite{Prodromou2012} The lid and in particular its L2 loop (residues 104-114) demonstrates significant variability both in the numerous crystal structures as well as in Molecular Dynamics (MD) simulations of the Hsp90 NTD.\cite{Colombo2008} Changes in the orientation or positioning, or in some cases partial or complete folding, are some of the L2 loop conformational differences related to the Hsp90 chaperon activity \cite{Pearl2016} and nucleotide binding (unbound, ATP-bound, ADP-bound, etc).\cite{Prodromou1997,Huai2005,Krukenberg2011,Rashid2020} 

Despite the importance of the flexibility of this loop, its mechanistic origin has not been
understood due to the large time and space scales involved.\cite{Pearl2016} Insights into intra-state protein dynamics are key for better understanding the transitions between the different HSP90 states and their eventual targeting for drug design purposes.

%------------------------------------------------------------------------%
\subsection{The ATP-lid and Hsp90 inhibition}
HSP90 is one of the most studied pharmacological targets with more than 13,800 publications during the last 40 years [source: PubMed] and more than 270 available structures in Protein Data Bank. In 2022, a first Hsp90 inhibitor, pimitespib, has been approved in Japan for patients with gastrointestinal stromal tumor. Most of the small molecules targeting Hsp90 bind to the ATP binding site in the NTD. By blocking ATP binding, these inhibitors prevent the release of the ATP-lid segment and the NTD dimerization, a critical initial step in the catalytic cycle of Hsp90 prior to its chaperoning activity.

Despite the undeniable importance of the ATP-lid that appears in different conformations in the available structures, it is not yet clear what is the driving force behind the transition between these different conformations. Interestingly, Colombo et al. using MD simulations of the NTD showed a spontaneous transition between the apo, i.e. nucleotide free, and the holo, i.e. nucleotide bound, conformations.\cite{Colombo2008} In other words, it is not the nucleotide binding/hydrolysis that determines Hsp90 conformations; instead Hsp90 exists in a conformational equilibrium between its different states and this equilibrium may or may not be shifted upon nucleotide or ligand binding.\cite{Colombo2008,Southworth2008,Hessling2009,Mickler2009,Zhang2015,Pearl2016,Rashid2020}  

%------------------------------------------------------------------------%
\subsection{\label{sec:states_def}Data mining HSP90 structural diversity}
The HSP90 NTD has been extensively studied and more than 270 times experimentally resolved. These structures, both with or without ligands, are available in the Protein Data Bank (PDB) and demonstrate some specific structural differences in the binding site L2 loop (residues 105-114). 

We used this plethora of structures to identify potential native states of the Hsp90 NTD. First, once all available structures of the NTD HSP90 were fetched from PDB (278 different conformations), they were aligned and necessary fixes were performed, i.e. numbering after sequence alignment, completion of partially resolved residues, number of chains per PDB, atom naming, tags etc. The protein structures were prepared with hydrogens being added and protonation being assigned using the Schr\"{o}dinger Protein Preparation Wizard~\cite{Sastry2013}. Then, the dihedral distribution of the L2 loop was used for the clustering of the structures. More precisely, for each conformation, the sine and cosine values of the dihedral angles $\Phi$ and $\Psi$ for residues 105 to 114 were computed. 

The resulting dataset, which contains 278 observations of 40 features (sine and cosine of the 2 dihedrals from each of the 10 residues), was used to perform clustering of the structures using the hierarchical clustering method implemented in \texttt{scikit-learn}. The metric used to perform clustering of the structures using the hierarchical clustering method implemented in \texttt{scikit-learn}. The metric used was the Euclidean distance in the (sin, cos) 40-dimensional space, so that the distance between two conformations $\alpha_1 = (\Phi_1^1, \Psi_1^1, \dots, \Phi_1^{10}, \Psi_1^{10})$ and $\alpha_2 = (\Phi_2^1, \Psi_2^1, \dots, \Phi_2^{10}, \Psi_2^{10})$ is given by

\begin{equation*}
\begin{split}
d(\alpha_1, \alpha_2)^2 & = \sum_{i=1}^{10}(\sin(\Phi_1^i)-\sin(\Phi_2^i))^2 +(\cos(\Phi_1^i)-cos(\Phi_2^i))^2  \\
                        & + (\sin(\Psi_1^i)-\sin(\Psi_2^i))^2 +(\cos(\Psi_1^i)-\cos(\Psi_2^i))^2.
\end{split}
\end{equation*}

\begin{figure}[hbt!]
\includegraphics[width=8cm]{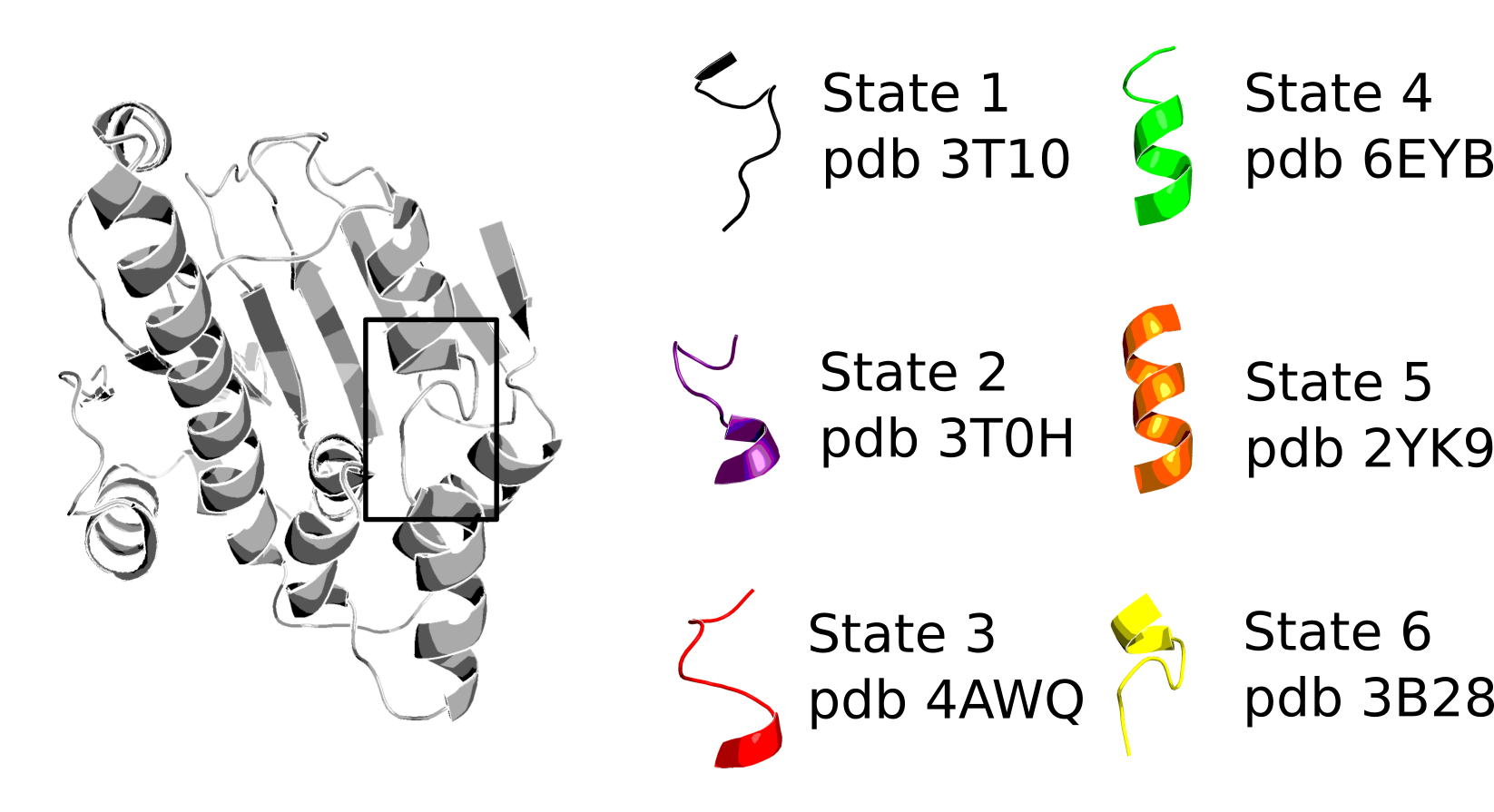}
\caption{The six representative Hsp90 states identified through clustering and the representative PDB IDs used for the present study.}
\label{fig:clustering}
\end{figure}

Six clusters were identified, indicating the existence of six key states (Fig.~\ref{fig:clustering}). These states can be differentiated by a loop/helix conformation formed by residues 105 to 114. States 1 and 2 exist in both apo/holo forms, while the rest of the states have been resolved in holo conformations only. A representative structure of each state was selected using as criterion the highest resolution. The exact PDB IDs are shown in Fig.~\ref{fig:clustering}. 

For the remainder of the work, each state will be referred to by its associated number. It should be also noted that the corresponding ligands have been removed from the holo structures before any simulations. All MD simulations described and analyzed in this work are thus of the unbound NTD. It is also important to recall that the states were defined based on the loop of interest alone (residue 105-114), and are thus not necessarily directly representative of the states of the whole Hsp90 protein. 

%------------------------------------------------------------------------%
\section{\label{sec:AE-CV} CV learning using an autoencoder}

Hsp90’s conformational cycle is a dynamic equilibrium between the open and closed states, passing through intermediate states, that are also accessible in the absence of nucleotide. It has been shown that the role of nucleotide is not to determine the conformation, but to lower the energy barriers between the states.\cite{Hessling2009,Mickler2009,Zhang2015,Pearl2016,Rashid2020} In the previous section, we demonstrated that the `known' Hsp90 conformational space can be clustered into six native states. As a next step, we aim at identifying potential transitions between these states through biased MD simulations using an autoencoder-learned collective variable space.

Most often, ML-constructed CVs are obtained via unsupervised learning (see however Refs.~\onlinecite{BRP20,BPP21} for notable exceptions). This is more than justified by the fact that supervised learning models rely on the knowledge of a label set $y$ associated with the dataset $X$. In the case of molecular dynamics, this label set assigns for example the index of a metastable state to each sample. Yet, the assumption that these states are known and distinguished makes supervised learning models non applicable in many cases. Nevertheless, in cases where conformational states are known and the samples can indeed be assigned to states, it seems essential to include this information to the learning of a collective variable, as this would help recover a CV which distinguishes the different conformational states.

%~~~~~~~~~~~~~~~~~~~~~~~~~~~~~~~~~~~~~~~~~~~~~~~~~~~~~~~~~~~~~~~~~~~~~~~~~~~~~~~~~~~~~~~~%
\subsection{Dataset generation}

The dataset used for the training of the autoencoder is composed of short MD simulations of each identified state (Fig.~\ref{fig:clustering}). For each of the 6 states, the protein structures were prepared in order to have the exact same number and naming of atoms (3258 atoms). Replicates of 20-ns MD trajectories were used as training set and were generated with Gromacs v5.1.3~\cite{GROMACS}. The Amber99-DISP all-atom force field~\cite{Robustelli2018} and the TIP4P model~\cite{TIP4P} were used to model the protein and water, respectively. All proteins were solvated in a cubic box with an 11 \AA~margin around the protein to ensure a sufficient minimum separation of the protein from its periodic images. Na+ counterions were randomly placed in the system to neutralize the total charge while a 0.15M of NaCl salt was also added to resemble the physiological conditions. One simulation per state representative structure was performed (PDB IDs reported in Fig. \ref{fig:clustering}). Before the production runs, a minimization and a 5-step restrained MD relaxation protocol was followed. First, all heavy atoms of the system were restrained using a force constant of $1000$~kJ/mol$\cdot$nm$^2$ for 100~ps. Then, only the protein atoms were restrained using a force constant of $1000$~kJ/mol$\cdot$nm$^2$ for 100~ps. Two  more 100~ps restrained MD with the same force constant were performed with restrains only on backbone and then only on $C_\alpha$ atoms. Finally, a flat bottomed restraint was applied on the $C_\alpha$ carbons, with a force constant of $500$~kJ/mol$\cdot$nm$^2$, starting at a distance larger than 2~\AA~from the reference structure and for 400~ps. The $C_\alpha$ atoms Root Mean Square Deviation (RMSD) during these last 400~ps of relaxation remained stable. The convergence of the production simulations to a stable conformation was evaluated using the total $C_\alpha$ carbon RMSD (Fig. S1). 

For each state, $10$~trajectories were independently sampled starting from the same configuration within that state and different initial velocities. We thus obtained $60$~trajectories of $20$~ns each. The final dataset consists of the concatenation of the short 20-ns trajectories sampling the six different conformational states, shown in (Fig.~\ref{fig:clustering}). All resulting configurations were aligned to the same reference structure, which is the representative conformation of State~1 (PDB ID: 3T10). It should be noted that States~4 and~5 converge even after a very short simulation time to the same conformation of the L2 loop and have therefore been merged for the rest of the study into the same state, i.e. State~5. The C$\alpha$ carbon coordinates were kept as input features, making the input dimension $D_{\rm B} = 3 \times 207 = 621$ coordinates. The dataset totals to $N_{\rm B} = 240,000$ points.

To ensure that the five states that have been identified using structural clustering are metastable even after the removal of ligands, we performed 200-ns unbiased MD starting from each of these states. RMSD and cluster centroid distance calculations show that the sampled simulations do not visit any of the other states, but possibly explore apparent substates (Figures S2 and S3).

%~~~~~~~~~~~~~~~~~~~~~~~~~~~~~~~~~~~~~~~~~~~~~~~~~~~~~~~~~~~~~~~~~~~~~~~~~~~~~~~~~~~~~~~~%
\subsection{Autoencoder architecture}
A symmetric AE architecture was used and only fully connected layers were considered. The number and size of the hidden layers determine the complexity of the model and thus of the learned collective variables. Generally, when layers are added, more complex representations can be modeled by the AE. However, it should be taken into account that more layers also require a larger dataset for training and can lead to overfitting. Moreover, from a practical point of view, the purpose of the autoencoder collective variable is to run biased sampling. The run time of the biased sampling simulation dramatically increases with more complex CVs.

To examine whether the addition of a hidden layer improves the learned model, we considered two AE architectures, with the size of the layers being chosen so as to gradually reduce the
dimensionality from input to bottleneck: 

\begin{itemize}
    \item $S_1$, one hidden layer between the input and bottleneck: \\
    Input (621) $\xrightarrow{}$ Hidden 1 (100) $\xrightarrow{}$ Bottleneck ($k$) $\xrightarrow{}$ Hidden 2 (100) $\xrightarrow{}$ Output (621).

    \item $S_2$, two hidden layers between the input and bottleneck: \\ 
    Input (621) $\xrightarrow{}$ Hidden 1 (150) $\xrightarrow{}$ Hidden 2 (40) $\xrightarrow{}$ Bottleneck ($k$) $\xrightarrow{}$ Hidden 3 (40) $\xrightarrow{}$ Hidden 4 (150) $\xrightarrow{}$ Output (621).
\end{itemize}

The numbers between brackets correspond to the size (i.e. number of neurons) of each layer. The parameter~$k$ is the dimensionality of the bottleneck layer (the final layer of the encoder, i.e. the CV dimension). We used different values of~$k$, ranging from~1 to~10, to identify the optimal CV dimension. For each value of~$k$, two AEs with structures $S_1$ and $S_2$ were trained. 

The autoencoders were constructed and trained using the \texttt{Keras} library\cite{Keras} in Python. All autoencoders are trained on~$75\%$ of the dataset, leaving~$25\%$ for validation. The learning rate used is $\eta=10^{-4}$ with \texttt{Adam} optimization. A batch size of~$1000$ samples was used and the training ran for a maximum of $1000$ epochs. Early stopping of the training is applied when the validation loss does not improve for~$40$ consecutive epochs to avoid overfitting. 

To compare the two AEs, we plot the evolution of their training and validation losses throughout training (Fig. S4). It can be observed that the training of the AEs with structure $S_1$ shows more stability, i.e. the evolution of the validation and training losses is approximately the same. Conversely, structure $S_2$ autoencoders (apart from the $k = 10$ autoencoder) overfit after $\sim$100 epochs. Overfitting is however already handled by the early stopping procedure and is thus not an issue. For each $k$, the optimal structure $S_1$ model, i.e. the model with the optimal validation loss, reaches approximately the same validation and training loss as the optimal $S_2$ model. In particular, for $k = 1$, it can be observed that $S_2$ seems to outperform $S_1$ on the training loss, but the corresponding validation loss evolution shows that this actually corresponds to the $S_2$ model overfitting. The only advantage of structure~$S_2$ is that the training finishes in fewer epochs. Nevertheless, since $S_2$ represents larger models, one training epoch takes longer to complete, meaning that this lower number of epochs does not necessarily translate to faster convergence in wall-clock time. More importantly, the most time consuming part of our procedure is by several orders of magnitude the biased simulations, which run approximately 20 times faster with an encoder CV from structure $S_1$ than one from structure $S_2$. We thus choose structure~$S_1$ for the autoencoders used in this work.

%~~~~~~~~~~~~~~~~~~~~~~~~~~~~~~~~~~~~~~~~~~~~~~~~~~~~~~~~~~~~~~~~~~~~~~~~~~~~~~~~~~~~~~~~%
\subsection{\label{sec:CV_dim}Choice of the CV dimensionality}

The optimal dimensionality of the CV, i.e. AE bottleneck layer size, for a certain system is defined as the minimal dimension of variables which are enough to represent a maximal portion of the patterns and features of the conformational space. Numerically, this can be translated as a trade-off between the subspace dimensionality and the amount of data variance covered by that subspace. In the case of a PCA for example, this means keeping the principal components (PCs) with the highest eigenvalues. More specifically, the optimal dimension is determined by some `elbow' in the scree plot (which corresponds to plotting the eigenvalues ranked in decreasing order). PCA therefore provides direct quantities to help determine the optimal dimension of a system: the eigenvalues. This is not the case for other models, such as the one used herein, i.e. autoencoders.

In order to choose the dimension of the bottleneck for our production calculations, we considered a set of 10 autoencoders with bottleneck layer dimensions ranging from~$k=1$ to~$k=10$ that were trained on our dataset using the architecture $S_1$ (see previous section). The corresponding training curves (loss optimization and validation loss evolution) were kept, as well as the last obtained model and the model which achieved the best validation loss. 

To compare these $10$ models, we plot their corresponding loss evolution and compare their final losses (Fig. S5). 

Based on our results, two CVs would be able to provide enough information, while the optimal dimension of the AE bottleneck is~5 (Fig. S5). The loss values do not significantly decrease as the bottleneck size $k$ increases, which means that increasing the CV dimensionality does not significantly improve the reconstructed output. The additional dimensions of the CV evidently learn directions that are of much lower variance compared to the initial 1-dimensional bottleneck. A possible explanation is that these directions correspond to learning noise in the training data. However, because the validation curves are similar to the training curves, this hypothesis can be discarded. Additionally, plotting the 1-dimensional bottleneck encoder CV over the training dataset shows that this direction alone is actually not able to differentiate between all the identified states of Hsp90. We therefore argue that despite their relatively small variance, the additional dimensions may still be of importance. Also, as our goal is to train the AE using a relatively small latent space, we select the dimensionality after which some fluctuations appear (possibly due to small differences in the loss function not showing up because of stochastic errors in the optimization procedure), i.e. $k^\star= 5$.  

The final autoencoder architecture used for the rest of our study is composed of four layers with the following sizes: Input (621) $\xrightarrow{}$ Hidden 1 (100) $\xrightarrow{}$ Bottleneck ($k=5$) $\xrightarrow{}$ Hidden~2 (100) $\xrightarrow{}$ Output (621). The activation function used for all
layers is hyperbolic tangent.

%\newpage
\begin{figure*}[htb!]
\includegraphics[width=0.95\textwidth]{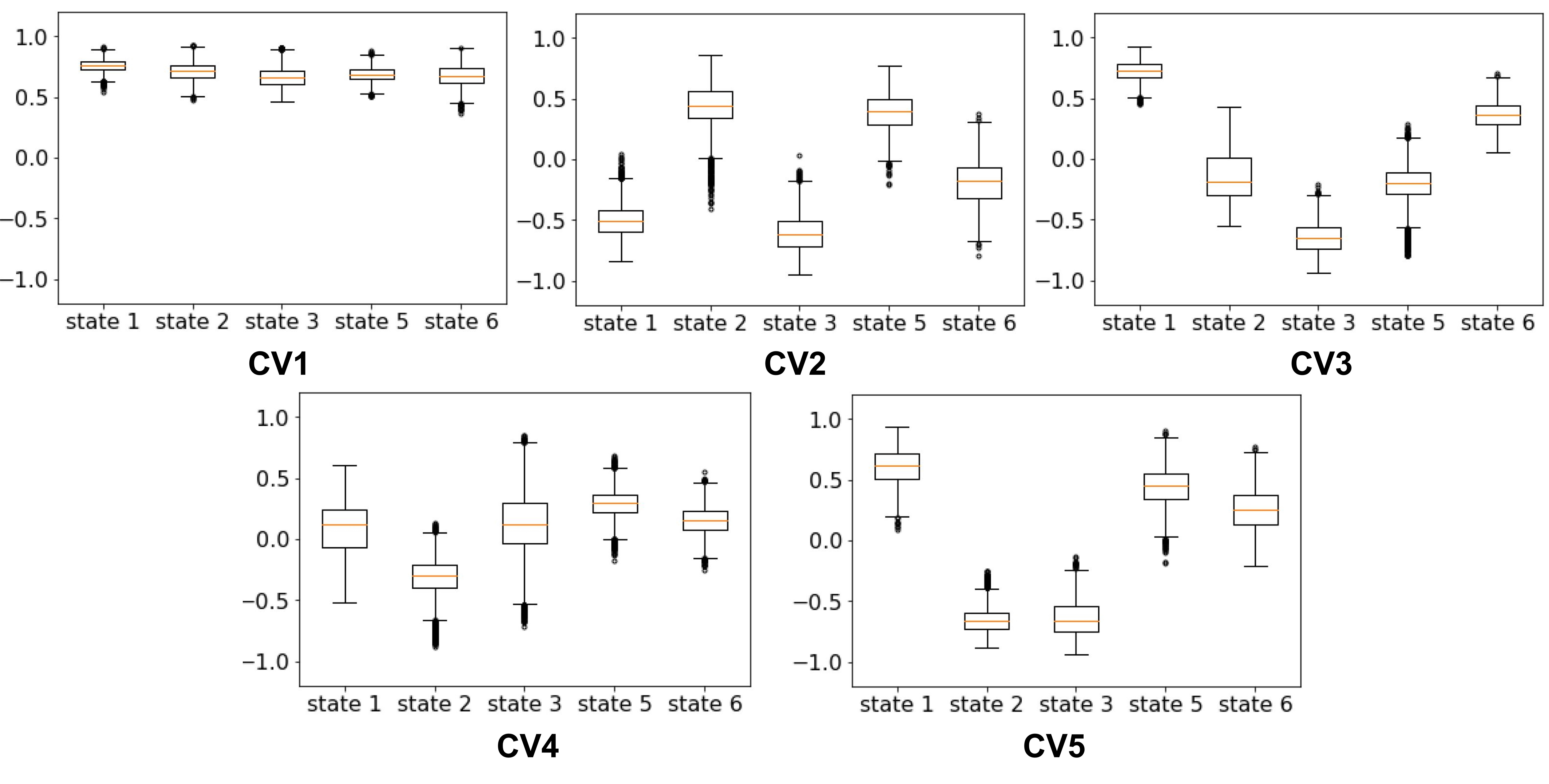}
\caption{Boxplots of the CV variation per state.}
\label{fig:CV_variation}
\end{figure*}
%~~~~~~~~~~~~~~~~~~~~~~~~~~~~~~~~~~~~~~~~~~~~~~~~~~~~~~~~~~~~~~~~~~~~~~~~~~~~~~~~~~~~~~~~%
\section{\label{sec:optimal_CV} Choosing the `optimal' collective variable for Hsp90}

Once the autoencoder is trained, the encoder is used as the learned collective variable. In theory, the 5-dimensional CV could be used directly. In practice however, the simultaneous biasing in 5 directions and estimation of the free energy over this 5-dimensional space is prohibitively expensive from a computational viewpoint. ABF typically requires a low CV dimension of 1 to maximum~3. Our aim here is to reduce the CV to a 2-dimensional one. We explore two approaches: selecting a 2-dimensional CV out of the 5-dimensional one (see Section~\ref{sec:2_out_of_5}, and then compare the quality of this selected CV to directly learning a 2-dimensional CV (see Section~\ref{sec:directly_2_dim}).

\subsection{Selecting a 2-dimensional CV}
\label{sec:2_out_of_5}

To minimize the dimensions of the CV to be used for biasing, we assessed the efficiency of each component of the CV at describing and differentiating the five identified states of Hsp90, always based on the L2 loop conformations. First, we plotted the values of each of the five CV coordinates, which we refer to by $CV_i$ for $1 \le i\le 5$. For this, we separate our training dataset into the 5 conformational states to observe how each coordinate of the CV varies from one state to another (Fig.~\ref{fig:CV_variation}). The results indicate that $CV_1$ cannot discriminate between the five identified states of Hsp90. Next, $CV_4$ takes comparable values over all states, and therefore does not provide a clear separation between states either. The coordinates $CV_1$ and $CV_4$ are therefore eliminated, as they fail to differentiate significantly between the five states and are thus not expected to be helpful for driving transitions among these states.

\begin{figure}[htbp!]
\includegraphics[width=0.37\textwidth]{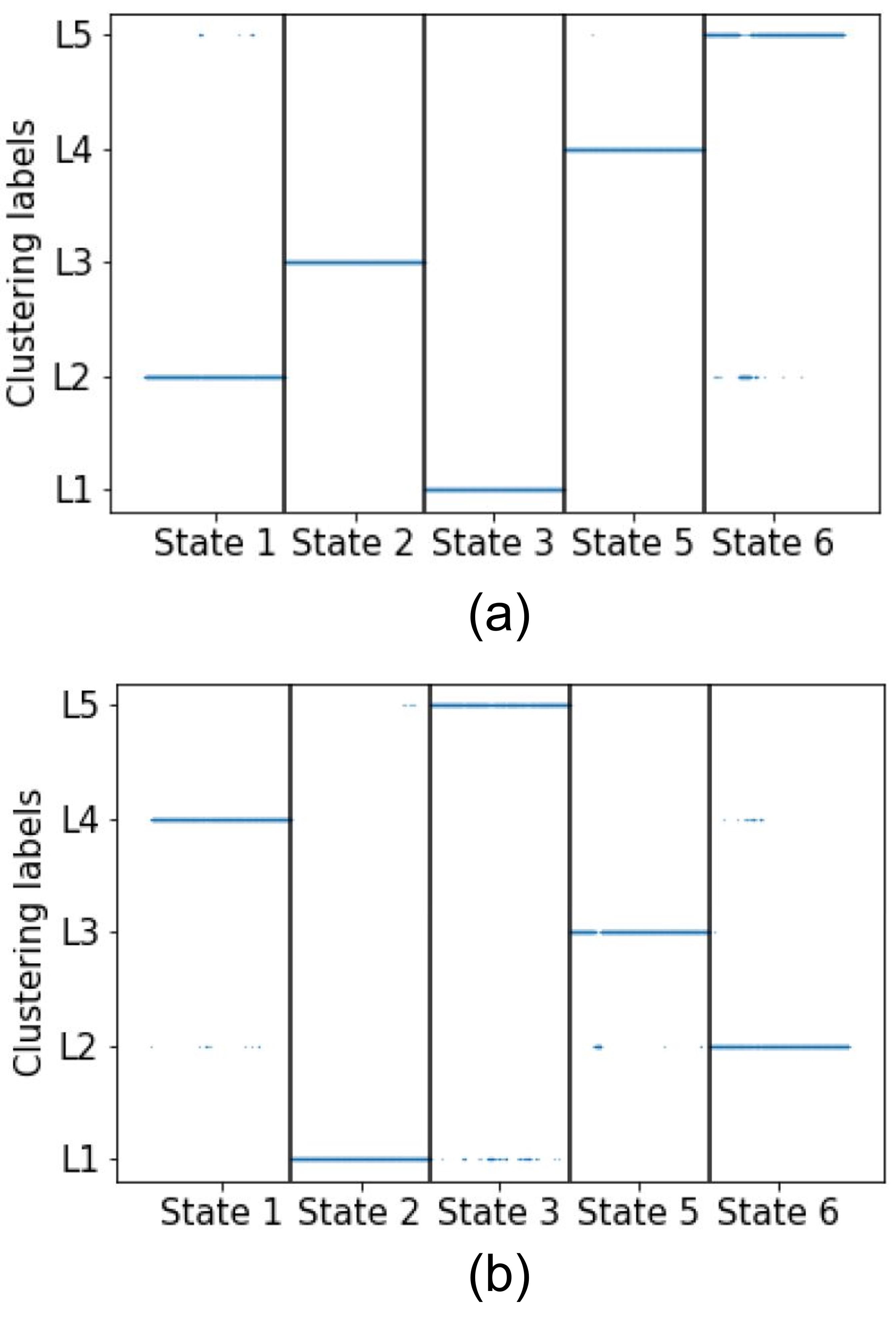}
\caption{Agglomerative clustering of the dataset comprising short MD simulations of the five different L2 loop states. The vertical axis corresponds to each cluster represented by its label number L1 to L5, while the horizontal axis corresponds to the conformational states defined in Sec.~\ref{sec:states_def}. Results obtained using (a) the 5-dimensional CV space and (b) the coordinates ($CV_3$;$CV_5$).}
\label{fig:Clustering_5D-2D}
\end{figure}

Next, to discriminate between the remaining three directions, i.e. $CV_2$, $CV_3$ and $CV_5$, we perform hierarchical clustering over each pair of CVs, and select the CV pair whose clustering best separates the five states. For this, we use the agglomerative clustering method, with Ward's minimum variance criterion to merge clusters. The method starts with as many clusters as the number of datapoints in our dataset, and successively merges clusters by minimizing the variance of the newly merged clusters. We performed the agglomerative clustering over the 5-dimensional CV space and over all three possible CV pairs, namely ($CV_2$;$CV_3$), ($CV_2$;$CV_5$) and ($CV_3$;$CV_5$). We then stop the hierarchical clustering at 5 clusters aiming at identifying the five Hsp90 L2 loop states identified previously. Interestingly, the five states are recovered both for the 5-dimensional CV and the combination of coordinates ($CV_3$;$CV_5$) (Fig. \ref{fig:Clustering_5D-2D}). Moreover, the dendrograms corresponding to the clustering indicate that clusters L2 and L5, i.e. States~1 and~6, are the most similar compared to the remaining states (Fig. S6). 

Based on this analysis, we can conclude that the combination of CV coordinates ($CV_3$;$CV_5$) is able to separate the 5 states with the highest accuracy, i.e. the lowest number of mislabeled points, almost equally well as the 5-dimensional CV. In Fig.~\ref{fig:CV_state}(a), we plot  the CV coordinates ($CV_3$;$CV_5$) over the dataset of short unbiased trajectories started from each state (the samples from each state are colored using a different color). The plot shows that these two coordinates indeed differentiate between the five L2 loop conformational states, making it a good choice for running a free energy biasing procedure. For the remainder of this work, we refer to the coordinate pair~$(CV_3,CV_5)$ as the autoencoder CV, but keep the indexing~$CV_3$ and~$CV_5$. 

\subsection{The effect of bottleneck dimensions}
\label{sec:directly_2_dim}
\begin{figure}[hb!]
\includegraphics[width=0.4\textwidth]{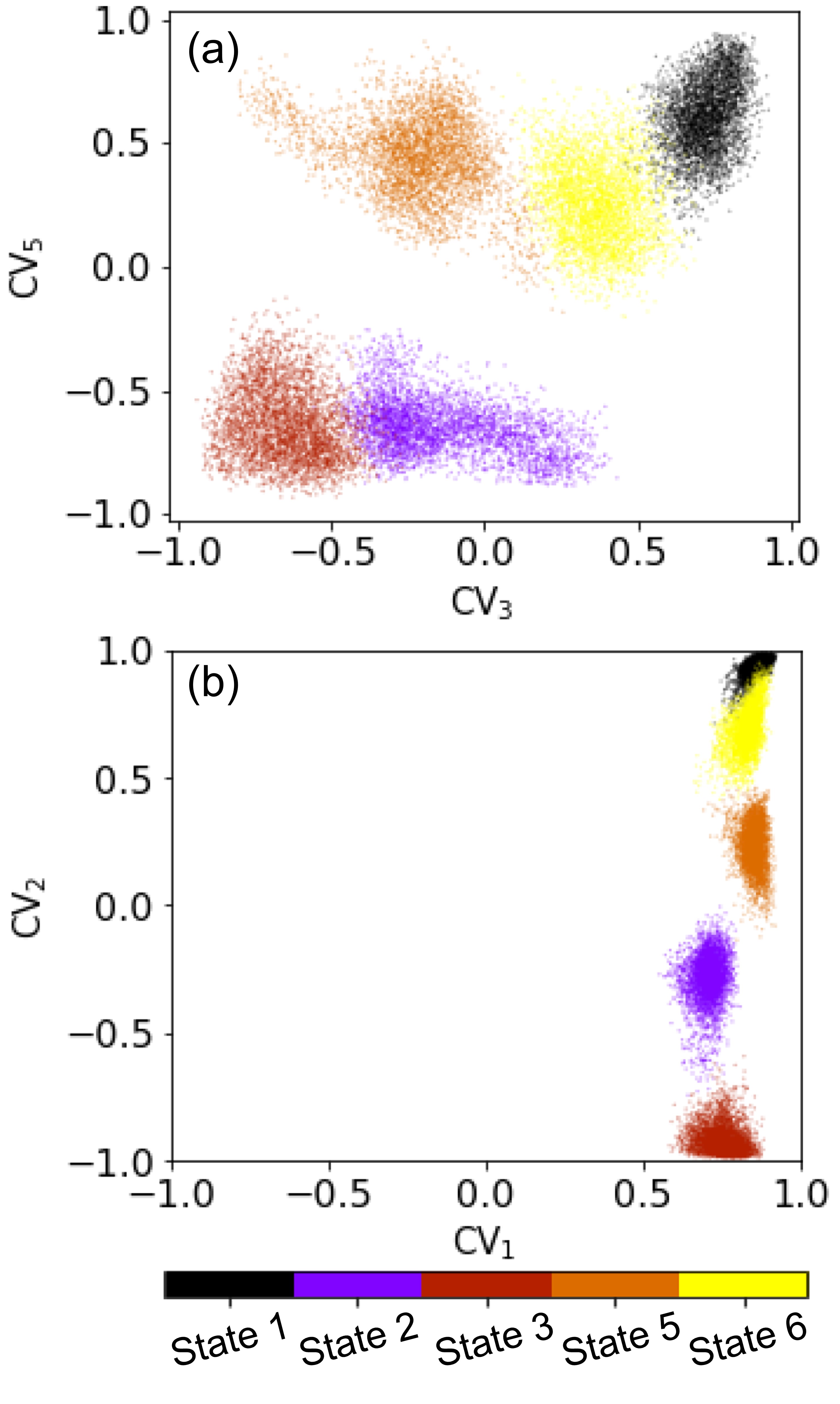}
\caption{Scatter plot of the autoencoder CV. Each point is colored according to its corresponding state. Top: Coordinates~3 and~5 of the 5-dimensional bottleneck autoencoder. Bottom: Coordinates~1 and~2 of the 2-dimensional bottleneck autoencoder.}
\label{fig:CV_state}
\end{figure}

In light of the results of the previous section, i.e. a 2-dimensional CV is able to separate the 5 different L2 loop states of Hsp90, and the discussion in section \ref{sec:CV_dim}, one might argue that it might not be necessary to use a bottleneck dimension of~$k^\star=5$ and make a selection of two coordinates; instead, a direct training of an AE with a bottleneck dimension of $k^\star=2$ could be sufficient. As an unsupervised model trained for reconstruction of the input, the autoencoder will have to learn some features that do not necessarily distinguish between the metastable states (as CV1 in Fig. \ref{fig:CV_variation} for instance), but may nonetheless be important for data reconstruction (e.g. features representing a large number of residues of the protein). This is especially true for large proteins whose various states are sometimes only distinguished by motions in relatively small functionally important regions. Training an autoencoder with a large bottleneck size $k^\star=5$ makes it possible to learn and separate features representing the actual motions of interest from the features representing high variance directions that are unrelated to these motions. Then, handpicking the $d < k$ bottleneck dimensions of interest ensure more efficiency for biasing, compared to directly learning a CV using a $d$-dimensional bottleneck autoencoder. To illustrate our point, we compare in Fig.~\ref{fig:CV_state} the CV obtained from the selection of ($CV_3$;$CV_5$) coordinates against a 2-dimensional bottleneck autoencoder. The first coordinate of the second CV fails to distinguish between any states, while the second coordinate does not make a clear separation between States~1 and~6. The CV learned by an AE with a 2-dimensional bottleneck is therefore of lower quality than the one extracted from a 5-dimensional bottleneck AE.

%------------------------------------------------------------------------%
\section{\label{sec:HSP90-dynamical} Exploring HSP90 transitions using AE-learned CVs and eABF}

%~~~~~~~~~~~~~~~~~~~~~~~~~~~~~~~~~~~~~~~~~~~~~~~~~~~~~~~~~~~~~~~~~~~~~~~~~~~~~~~~~~~~~~~~%
\subsection{Technical information}

All the biased MD simulations were performed using OpenMM\cite{OpenMM} under its Python API. The starting conformation for each state is the last conformation after the position restraint MD. As previously discussed, the Amber99SB-ILDN forcefield and the TIP4P water model were used, with a cubic box of 12~\AA~buffer around the protein. Simulations were run in the NVT ensemble, with a Langevin integrator, a collision rate of 1~ps$^{-1}$, a timestep of 2~fs, and temperature $T = 300$~K. The extended system adaptive biasing force algorithm implemented in \texttt{PLUMED}\cite{plumed} was used for biasing. Each coordinate of the CV was discretized into~50 bins. All other parameters of the ABF algorithm were kept to their default values in \texttt{PLUMED}.

MD trajectories generated in this work are initially analyzed to ensure that the system remains stable throughout the simulation. All RMSD computations were done using the {\tt mdtraj} library\cite{MDTraj} in Python. Only the $C_\alpha$ carbons were used in the RMSD computations. For each trajectory, the RMSD was computed with respect to each of the 5 states, over the whole protein, and over the 10 residues forming the L2 loop of interest (105-114). Moreover, the dihedral angles of the residues forming the L2 loop were calculated post simulation using {\tt mdtraj}. The dihedral analysis uses the initial dihedral clustering model which was used to define the states (Section~\ref{sec:states_def}). The clustering distances, i.e. the distance of each conformation to the centers of the identified clusters, were used on the unbiased trajectories to delimit the states/clusters from a dynamical (rather than crystal structure based) viewpoint. This makes it possible to detect transitions in the biased trajectories by computing these same distances. As an additional analysis step, the hydrogen bonds formed within the L2 loop were computed for each state using {\tt VMD},\cite{VMD} with threshold values 25$^\circ$ for the acceptor-donor-hydrogen angle and 0.35~nm for the donor-acceptor distance. More precisely, the short unbiased trajectories are used to compute the frequency of the hydrogen bonds for each state. These computations were used to determine a set of the hydrogen bonds specific to each state, making it possible to define an H-bond-based fingerprint for each state, which can then be used to detect (or rather confirm) transitions in subsequent biased simulations.

%~~~~~~~~~~~~~~~~~~~~~~~~~~~~~~~~~~~~~~~~~~~~~~~~~~~~~~~~~~~~~~~~~~~~~~~~~~~~~~~~~~~~~~~~%
\subsection{\label{sec:transitions} Transition from State 1 to State 3}

After the selection of the best 2-dimensional candidate CV for Hsp90, i.e. ($CV_3$;$CV_5$), we initiated a series of biased simulations starting from one state and trying to explore transitions to other states, either already identified (Fig. \ref{fig:clustering}) or unexplored. 

\begin{figure}[ht!]
\includegraphics[width=0.4\textwidth]{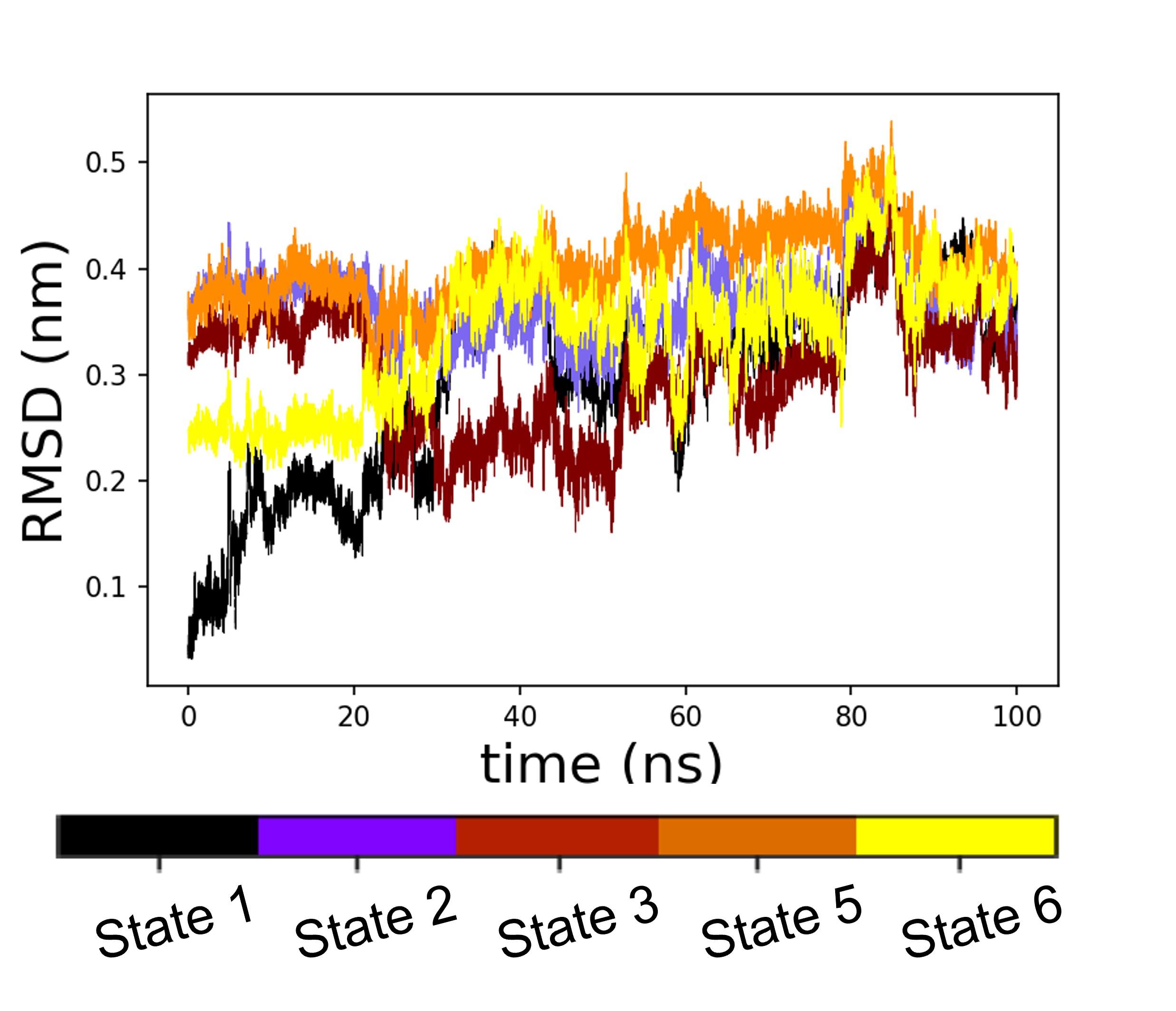}
\caption{RMSD values of the biased trajectory with respect to each state.}
\label{fig:rmsd_ex1}
\end{figure}

To showcase the ability of the autoencoder CV to facilitate the transitions between different states, we selected the example of a 100~ns biased MD simulation starting from State~1. Both coordinates of the autoencoder CV, i.e. biasing CV, span the whole range of the $[-1, 1]$ space (Fig. S7). This indicates that there is indeed a transition from State~1 to another state. To identify the visited states, the RMSD of the biased trajectory is calculated with respect to the initial conformation (i.e. energy minimized, and position restrained conformation) of each of the 5 states, using only the $C_\alpha$ carbons of the L2 loop. As expected, the minimal RMSD is at the beginning of the simulation obtained with respect to State~1, i.e. the starting state. Interestingly, at approximately 30~ns of biased MD and for 25~ns, the minimal RMSD shifts to the one with respect to State~3 (Fig. \ref{fig:rmsd_ex1}). After 55~ns the RMSD is increasing and possibly another previously not known state was visited.

The clustering model trained to obtain the six original states (Section \ref{sec:states_def}) was also used to characterize the potential transitions to different states. More precisely, the distance of each frame of the biased trajectory to each of the 5 cluster centers corresponding to States~1, 2, 3, 5 and 6 was calculated. Recall that distances to cluster 4 are not plotted as we have already discarded this state after merging it with state 5. As a point of comparison, the same distances are plotted for the unbiased dataset of trajectories started from each state (Fig.~S6). The results obtained on the unbiased trajectories demonstrate that to label a conformation as State~3, its distance to the centroid of the corresponding cluster 3 should range between~2.8 and~4, and its distance to the remaining centroids should be higher than~4. The same applies for all other states except State~5, for which the distances are smaller. The results obtained on the biased trajectory show that the computed distances are compatible with a transition to State~3 between 30~ns and 55~ns, i.e. approximately the same time frame for which a transition was identified using the RMSD (Fig. \ref{fig:rmsd_ex1}). We also note a small time frame around 25~ns where the conformations are closest to the centroid of State~5 than to any other state. These sampled points are thus automatically classified as State~5 configurations. However, their distances to the centroid~5, ranging between~3 and~4, are actually higher than the typical distance observed in State~5 unbiased trajectories, and which ranges between~1 and~2. This indicates that these conformations probably do not belong to State~5 (or any of the other states determined in the beginning of this work). We note that these `pseudo-state 5' conformations were sampled just before the time when the system transitions to State~3, and may therefore belong to a transient state between~1 and~3. Importantly, this misclassification emphasizes the importance of looking into the clustering distances, rather than simply trusting the cluster assignments.

\begin{figure}[ht!]
\includegraphics[width=0.37\textwidth]{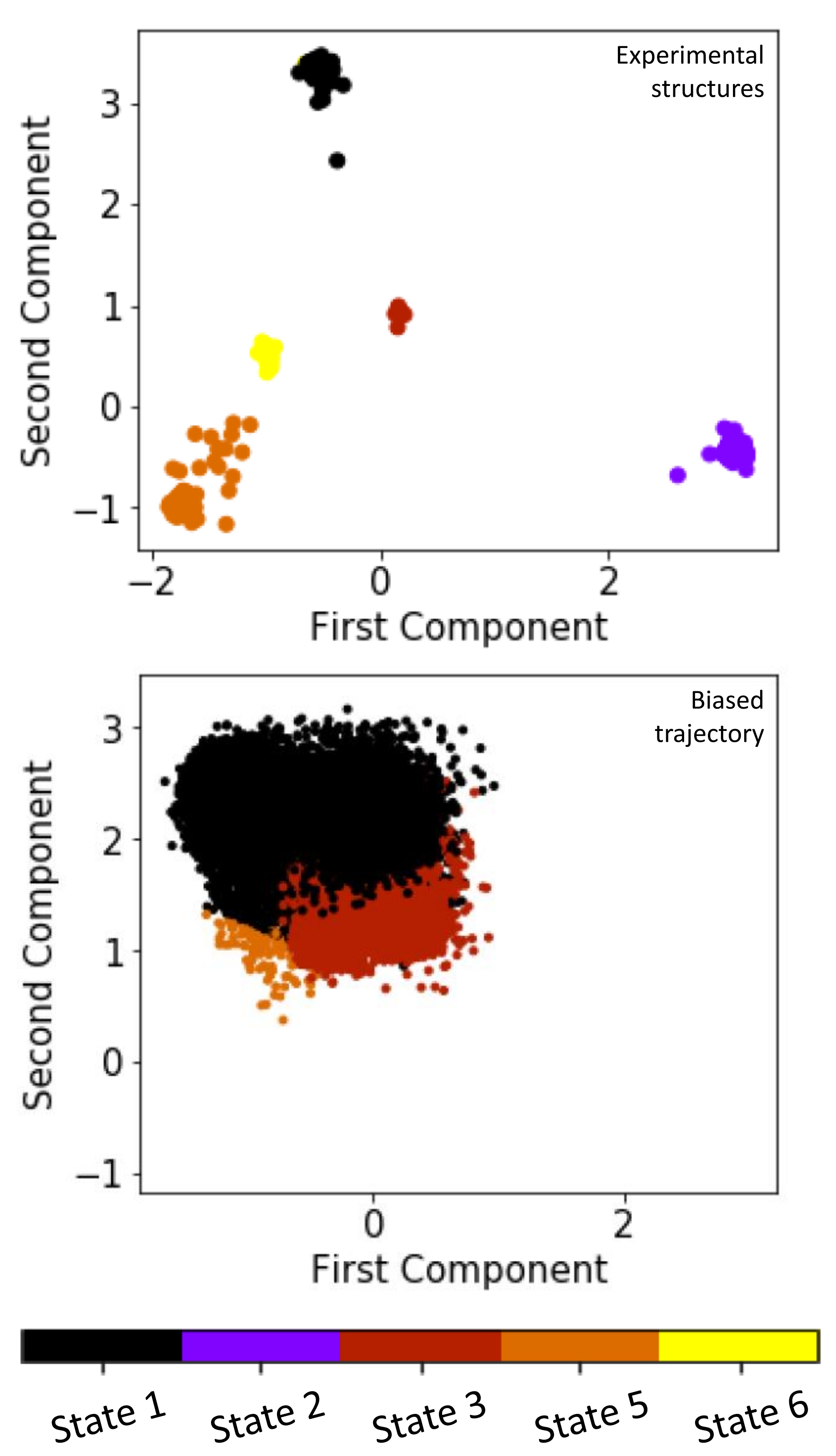}
\caption{Dihedral PCA over the dihedral angles of the L2 loop (residues~105 to~114). Top: Dihedral PCA of the experimentally resolved structures. Points are colored according to their states defined in Section \ref{sec:states_def}. Bottom: Dihedral PCA projection on the biased trajectory. Points are colored according to the cluster assigned using the dihedral clustering model (Fig. 1).}
\label{fig:dih_PCA}
\end{figure}

Dihedral PCA was also employed to identify possible transitions. Initially, we applied dihedral PCA on the 278 crystal structures already used for the definition of states to reduce the 20 dihedrals of the L2 loop into 2 principal components (Fig. \ref{fig:dih_PCA}, top). The obtained 2-dimensional space of the two principal components is able to discriminate the various crystal structures belonging to the different states and are colored accordingly. After applying this PCA projection to the biased trajectory, it can be inferred that what was previously assigned as State~5 (or pseudo-state 5) is most probably a new transient state between States~1 and~3 (Fig.~\ref{fig:dih_PCA}, bottom). 

\begin{figure*}[htb!]
\includegraphics[width=0.8\textwidth]{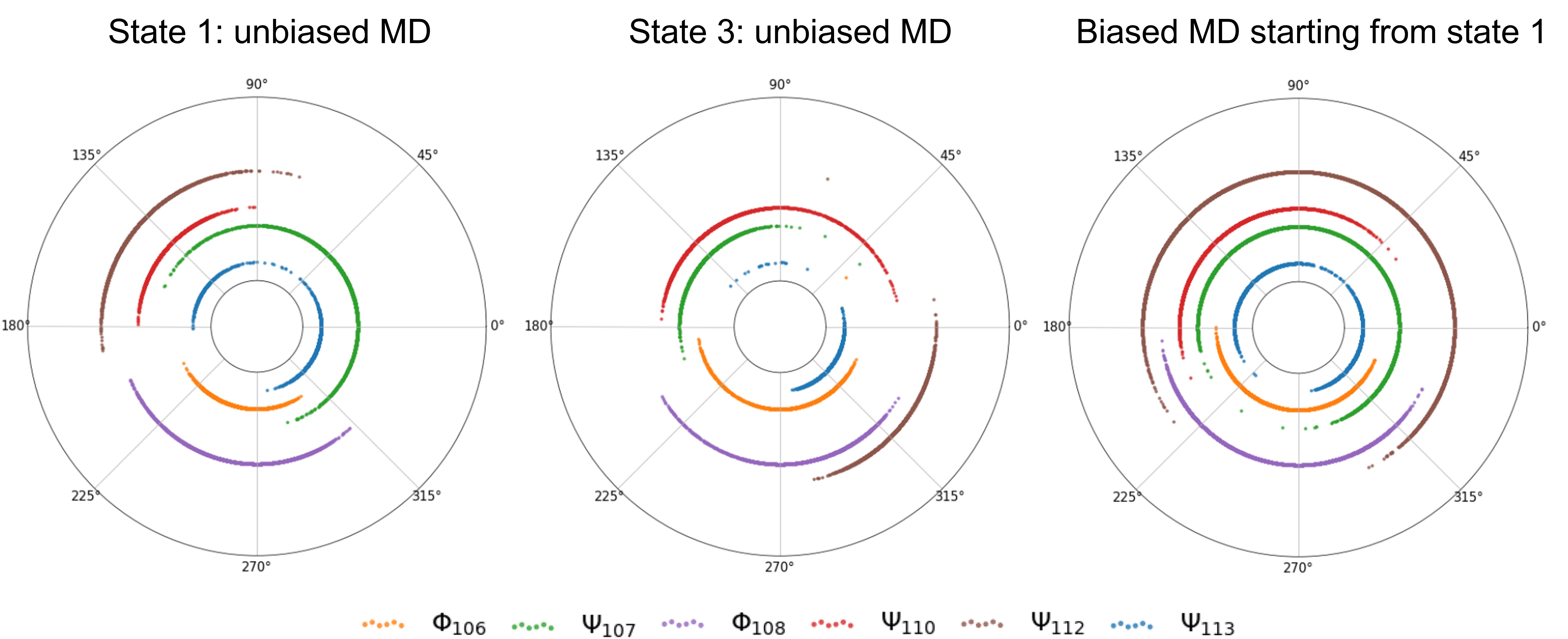}
\caption{Distribution of the six LASSO selected dihedral angles for (a) State~1, (b) State~3, and (c) the biased trajectory. For each of the two states the distribution has been calculated using the corresponding dataset of unbiased MD simulations.}
\label{fig:dih_angles}
\end{figure*}

To be able to further characterise the conformational differences between the different states, we defined a `fingerprint' of each state computed as the range of values taken by each dihedral angle in the L2 loop for each one of the states. The L2 loop comprises 10 residues or 20 dihedral angles and thus the obtained fingerprint can be hard to analyze. To guide the selection of dihedral angles sufficient to separate the different states, we trained a standard logistic regression model with $\ell_1$-norm regularization on the 278 crystal structures. This so-called LASSO regularization ensures sparsity in the trained model, so that only relevant and non redundant features, in our case dihedral angles, are kept. The obtained model selected six dihedral angles (Fig.~\ref{fig:dih_angles}). Here, we seek to define States~1 and~3 through the values of their dihedral angles.

 Fig.~\ref{fig:dih_angles} shows the ranges of these 6 dihedral angle values on States~1 and~3, as well as the values visited during the biased simulation of our example. The ranges of the dihedral angles sampled during the biased simulation can be seen as an interpolation between the typical values identified for States~1 and~3, inferring again that a transition between these states has most probably occurred. This is for example illustrated by the values taken by the dihedral angle~$\Psi_{112}$, which is colored in brown in Fig.~\ref{fig:dih_angles}. The angle~$\Psi_{112}$ ranges between $90^\circ$ and $190^\circ$ for State~1, and between~$-80^\circ$ and~$0^\circ$ for State~3; while its range during the biased simulation is~$[-70^\circ,200^\circ]$, thus interpolating the values observed in both States~1 and~3.

\begin{figure*}[htb!]
\includegraphics[width=0.8\textwidth]{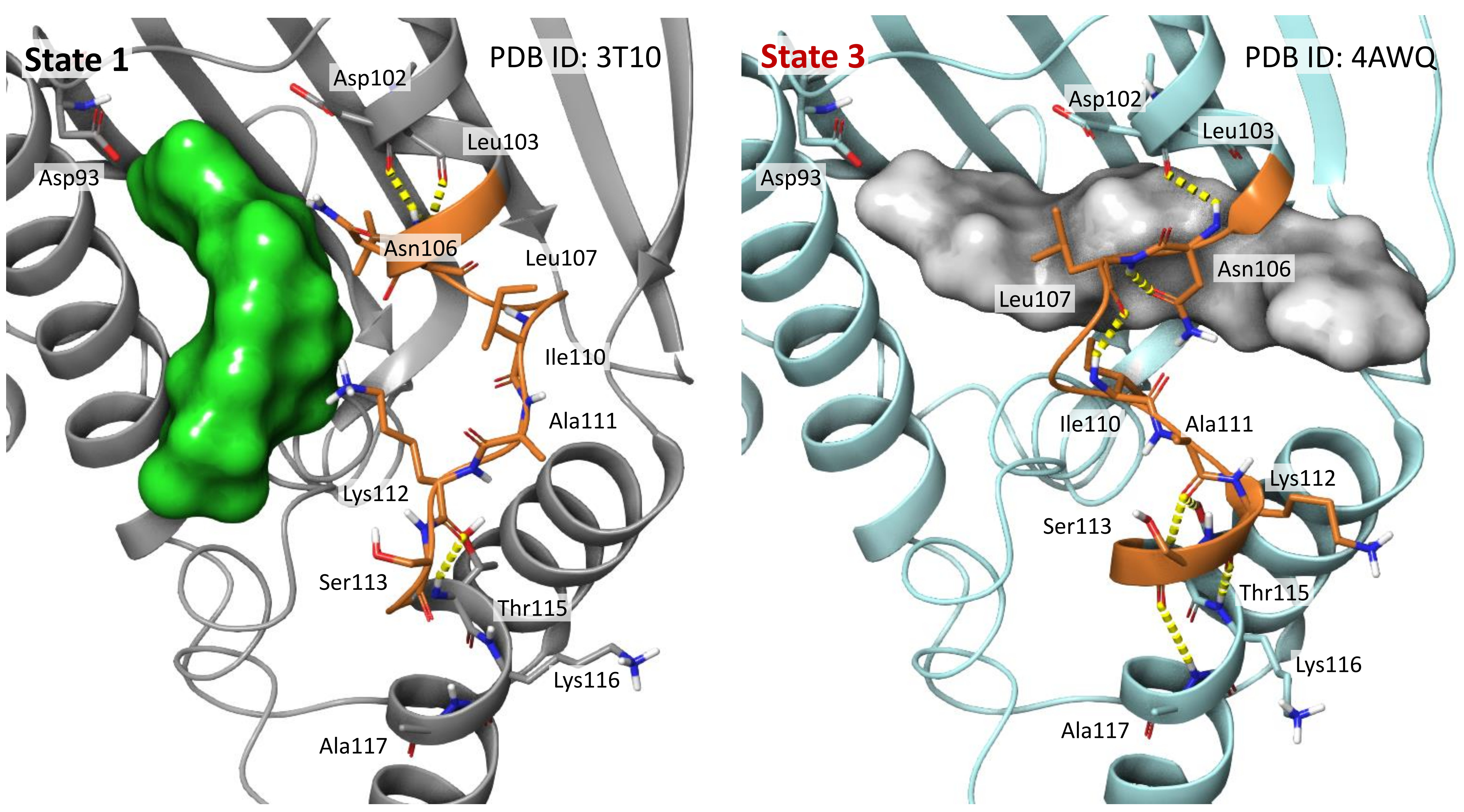}
\caption{Hsp90 NTD ATP binding site for State~1 (left, represented by PDB ID:3T10) and State~3 (right, represented by PDB ID: 4AWQ). The protein is shown in ribbon representation colored in grey for State~1, cyan for State~3, while the ATP-lid is colored in orange. The unique per state high frequency hydrogen bonds are explicitly shown (yellow dashed lines) while participating amino acids are shown in licorice representation. The positioning of the ligands in the initial structures is shown in green and grey surface for State~1 and State~3, respectively, but have been removed during the biased MD simulations.}
\label{fig:Hbonds}
\end{figure*}

To further characterize the different states, using the ten 20-ns trajectories of the initial dataset, we identified all the backbone hydrogen bonds for residues 100 to 120. A total of 70 hydrogen bonds appear in at least one of the 5 states and in at least one of the 10 short trajectories. For each hydrogen bond~$h$ and each conformational state~$i$, we determine a mean~$m_{h;i}$ and standard deviation~$\sigma_{h;i}$ of the hydrogen bond’s occurrence over the total sample of the specific state. Then, for each hydrogen bond $h$, we compute the overall mean $m_h$ and standard deviation $\sigma_h$ from the means $m_{h;i}$. To rule out hydrogen bonds that appear similarly often in all states or that appear rarely, we keep only hydrogen bonds with high enough values of $m_h$ and $\sigma_h$, i.e., $m_h > 4\%$ and $\sigma_h > \frac{m_h}{4}$. The final number of hydrogen bonds with high occurrence and variation among the states are 43. If we now keep only hydrogen bonds with high occurrence per state, i.e. $m_{h;i}> 10\%$ for at least one state, we have 19 hydrogen bonds left (Fig. S8). These high occurrence, high variance hydrogen bonds can be used to characterize each state. Focusing on States 1 and 3, there are three hydrogen bonds that appear only in State 1, namely Thr115-Lys112, Leu107-Leu103, and Asn106-Asp102 and four hydrogen bonds that appear only in State 3, namely L116-L112, Thr115-Ala111, Ile110-Leu107, and Ala117-Ser113 (highlighted in Fig. S9). In Figure~\ref{fig:Hbonds}, we have included the two starting structures for States 1 and 3 along with the hydrogen bonds identified as stable and unique for each state.

To examine the transition from State~1 to State~3 during the 100-ns biased simulation, we monitor the above-mentioned unique-per-state hydrogen bonds. Interestingly, three out of four hydrogen bonds that characterize State 3, namely Lys116-Lys112, Ile110-Leu107 and Ala117-Ser113, although not present at the beginning of the simulation (Fig.~\ref{fig:Hbonds}), gradually appear and coexist at approximately 48~ns and for 10~ns (Fig.~S9, (a), (b) and (c); the fourth hydrogen bond, i.e. Thr115-Ala111, appears for less than 1~ns and is thus not reported). On top of that, the three hydrogen bonds that exist only in State 1 and not in State 3 are broken during the biased simulation. Taken together, the formation of the three 'State 3' hydrogen bonds and the break of the `State 1' hydrogen bonds indicate a potential transition to a State 3-like conformation in agreement with the previous analyses. The hydrogen bond analysis demonstrates that on top of more holistic structural elements of the protein, e.g. $C_\alpha$ RMSD (Fig. \ref{fig:rmsd_ex1}) or PCA (Fig. \ref{fig:dih_PCA}), subtle differences in the vicinity of the binding site and in particular on the ATP-lid, occurring during the binding of different ligands, can be captured during a short biased simulation in the absence of ligand. This is of particular interest since, using our AE-ABF protocol and starting from a state that exists both in apo and holo forms, i.e. State 1, Hsp90-NTD transitions to a state where the ATP-lid is more structured and exists most probably only in a holo form (based on the existing experimentally resolved structures). 

%~~~~~~~~~~~~~~~~~~~~~~~~~~~~~~~~~~~~~~~~~~~~~~~~~~~~~~~~~~~~~~~~~~~~~~~~~~~~~~~~~~~~~~~~%
\subsection{\label{sec:Other_transitions} Other observed transitions}

To ensure that the results presented in Section~\ref{sec:transitions} are not specific to the simulation at hand (ABF starting from State~1), additional biased simulations were run. In particular, we ran ABF using the same 2D-CV, i.e., ($CV_3$;$CV_5$), starting from States 1, 3 and 5 and extending the sampling time to 200 ns. The distance from the centroids representing each state and dihedral PCA confirm the transition from State~1 to State~3 and an eventual transition to State~5 (Fig.~S10 (a) and (b)). Moreover, as shown in Fig.~S10 (c) and (d), starting from State 3, we observe an intermediate transition to State~6 before the system finally transitions to State~5 (there are indications that State~2 is also transiently visited). Finally, analysis of the biased simulation of State~5 indicates a transition to State~1 (Fig.~S10 (e) and (f)). 

The above-mentioned transitions between States~1, 3, and~5, and occasional visits to States~2 and~6, can provide information about the potential mechanistic effect of ligand binding. For example, the binding of geldanamycin,\cite{Roe1999} a natural product with Hsp90 inhibitory action, leads to an ATP-lid conformation of Hsp90 similar to the conformation seen in the isolated apo human Hsp90 NTD (State~1).\cite{Stebbins1997} This conformation on the other hand is different from the one observed upon nucleotide binding, where hydrophobic surfaces are accessible for NTD dimerization (State 2). Based on Fig.~S10, no transitions between States~1 and~2 are observed and thus this supports further the argument that an inhibitor, geldanamycin in this case, may stabilize the NTD in an apo-like conformation, or allosterically shift the conformation of Hsp90 to a state that is incapable of client protein binding. In this context, the presented framework can be used to identify potential transitions (or lack thereof) between biologically relevant states.  

%~~~~~~~~~~~~~~~~~~~~~~~~~~~~~~~~~~~~~~~~~~~~~~~~~~~~~~~~~~~~~~~~~~~~~~~~~~~~~~~~~~~~~~~~%
\section{Discussion and next steps}

Let us first emphasize that the methodology proposed in this work is very general and could be applied in principle to many other systems: starting from known metastable configurations from the PDB, augmented with short molecular dynamics simulations to explore the associated metastable basins, we were able by using autoencoders to build collective variables which are sufficiently good to, first, distinguish between the metastable states, and, second, enable transitions between those states when used in a free energy adaptive biasing method. 
Metaphorically, we could say that, at least on the test case considered in this work, autoencoders were able to shed light on possible pathways between already lightened basins. Let us mention that we have also used autoencoders in an even less supervised setting, namely starting from only one metastable basin and progressively discovering other basins to refine the collective variables, see the Free-Energy Biasing and Iterative Learning with AutoEncoders (FEBILAE) method presented in Ref.~\onlinecite{belkacemi2021}.
Autoencoders thus seem to be efficient tools to build relevant collective variables for molecular systems, in particular for biological applications.
Of course, the method should be further tested to assess its performance, which
 may also depend on the machine learning and molecular setup, including the autoencoder structure and training parameters, and the MD simulation parameters, in particular the choice of the force field.

Coming back more precisely to the results obtained on Hsp90, the obtained states are visually different from each other, and display some level of metastability as showcased by the RMSD evolution of unbiased simulations. The autoencoder trained on the dataset of short unbiased simulations allows the construction of a low dimensional CV able to clearly differentiate between the conformational states. Additionally, the results obtained on the 100~ns biased simulation showcase that this autoencoder CV can be used to efficiently sample transition paths among at least two states. Running biased simulations using more classical choices of CVs, such as some dihedral angles of the L2 loop, did not yield such transitions. Training the autoencoder on all five states was central to obtaining this CV. The entire analysis is in fact based on the definition of the five states, which was done using the clustering of the experimentally-determined structures. The results are thus all guided by this classification choice. While we argue that our classification of the states is well defined, many other choices could have been applied, for example by basing the clustering on the binding site itself rather than on the L2 loop adjacent to it.

The obtained states differ by the conformation of the L2 loop, originally due to their binding (or lack thereof) with different ligands. We chose in this work to simulate Hsp90 unbound, i.e. by removing the ligand from the bound states and applying restrained MD to stabilize the obtained systems. In Fig.~S7, we report the $C_\alpha$ RMSD of 200~ns unbiased simulations of the different states. These states are shown to be stable as they do not achieve any transitions between the identified states over the time frame of 200~ns. This indicates that these conformations do exist within the apo form of Hsp90. Importantly, this is in agreement with other studies of the NTD, which observe that the apo state visits various conformations of the NTD lid~\cite{Colombo2008,Amaral2017}. While the stability of each of these states for the apo structure should be further examined using more MD simulations, we can conclude in the context of the present study that the five separate conformational states of the L2 loop can all be visited in the apo form, and transitions among these conformations have been achieved by biased sampling of the autoencoder CV. We emphasize again that biased sampling applied with more traditional CVs, namely a selection of the dihedral angles of the loop of interest, did not yield any transitions. 

To the best of our knowledge, enhanced sampling of the Hsp90 NTD alone in apo form has not been previously performed. Unbiased sampling of the apo NTD shows some structural motions of the L2 loop, but no clear transitions (e.g. between the helix conformation of State~5 and the loop conformations of other states) is achieved in the $\mu$s timescale.\cite{Amaral2017} Several studies report enhanced sampling simulations of the bound NTD (taken separately or within the whole dimer), where the collective variable used is often the distance between the NTD and the ligand\cite{Kawaguchi2013,Kokh2018,Ngo2019,Bianciotto2021}. These simulations often aim at computing the binding affinity of the ligands to the NTD and consist of short dissociation trajectories to compare a wide array of inhibitors. Some of these studies do differentiate between conformations of the L2 loop of interest,\cite{Nunes2020} by, e.g. distinguishing between ligands which bind to the various conformations (loop or helix). However, the aim of these studies is not to drive transitions among these conformations, and structural motions of the L2 loop during the dissociation MD simulations are not reported/observed. In Ref.~\onlinecite{Simunovic2012}, Umbrella Sampling of the whole Hsp90 dimer, using the angle formed by the two monomers, is performed for the apo structure, where the calculated free energy surface shows the presence of two low energy states corresponding to the stretched and compact conformations, but the authors report no corresponding important motions in the NTD site itself.

A natural next step for the present study is to compute the free energy landscape of the 2-dimensional autoencoder CV. Because the identified states can somewhat be linked to apo or holo conformations of Hsp90, calculating the free energy surface could provide insights into the ligand binding mechanism. Additionally, while we argue that our choice of state definition, based on the clustering of a large number of crystal structures, is adequate, it is not the only way to obtain the initial states. Free energy calculations can help redefine the conformational states of Hsp90, and possibly discover new states in the local minima of the energy landscape, which were not included in the present analysis.

Finally, another possibility is to forgo state definitions altogether and to use only one conformation of Hsp90 to build the learning dataset. Such unbiased simulations are expected to be trapped within the initial state or only visit conformations close to it, and so iterative algorithms such as FEBILAE\cite{belkacemi2021} or other methods\cite{mesa1,mesa2,rave,rave2019} could be used. Nevertheless, assessing the efficiency of such iterative protocols method on a large biological system like Hsp90 presents an interesting challenge.

\section*{Supplementary Material}
See the supplementary material for additional analyses not shown in the main text. In particular, we have included RMSD plots of the traning set, training loss convergence, clustering information, the collective variable evolution, a hydrogen bond analysss, and a dihedral angles analysis of some additional biased simulations.

\begin{acknowledgments}
The work of T.L. and G.S. is partially funded by the European Research Council (ERC) under the European Union’s Horizon 2020 research and innovation program (grant agreement No. 810367). 
\end{acknowledgments}

\section*{Data Availability Statement}
The data that support the findings of this study are available from Sanofi R$\&$D and ENPC. Restrictions apply to the availability of these data, which were used under license for this study. Data are available from the authors upon reasonable request and with the permission of Sanofi R$\&$D and ENPC.

\nocite{*}

\end{document}

\maketitle

\begin{figure}
\centering
\includegraphics[width=0.95\textwidth]{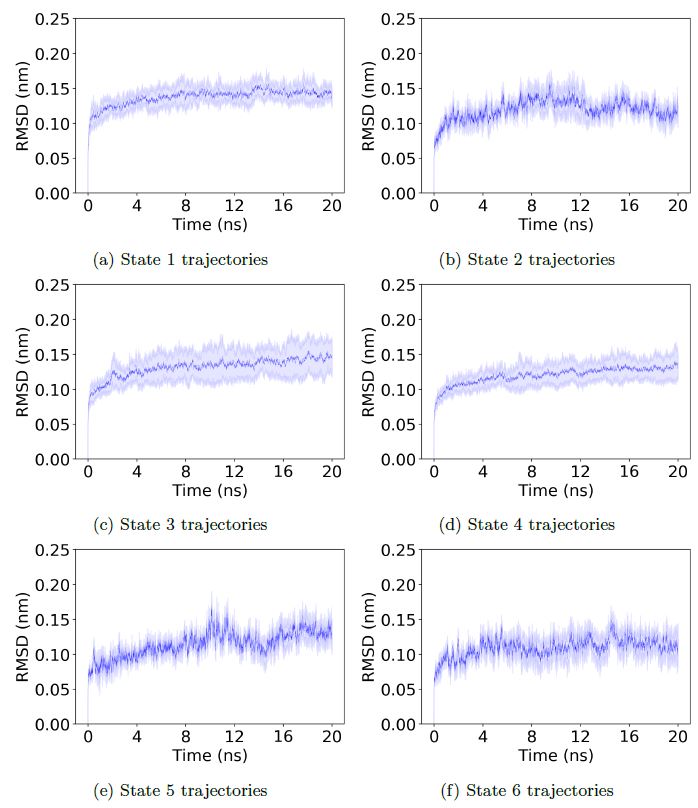}
\caption{Alpha carbon RMSDs mean and standard deviation over the 10 trajectories calculated for each state.}
\end{figure}

\pagebreak
\newpage

\begin{figure}[h!]
\centering
\includegraphics[width=0.95\textwidth]{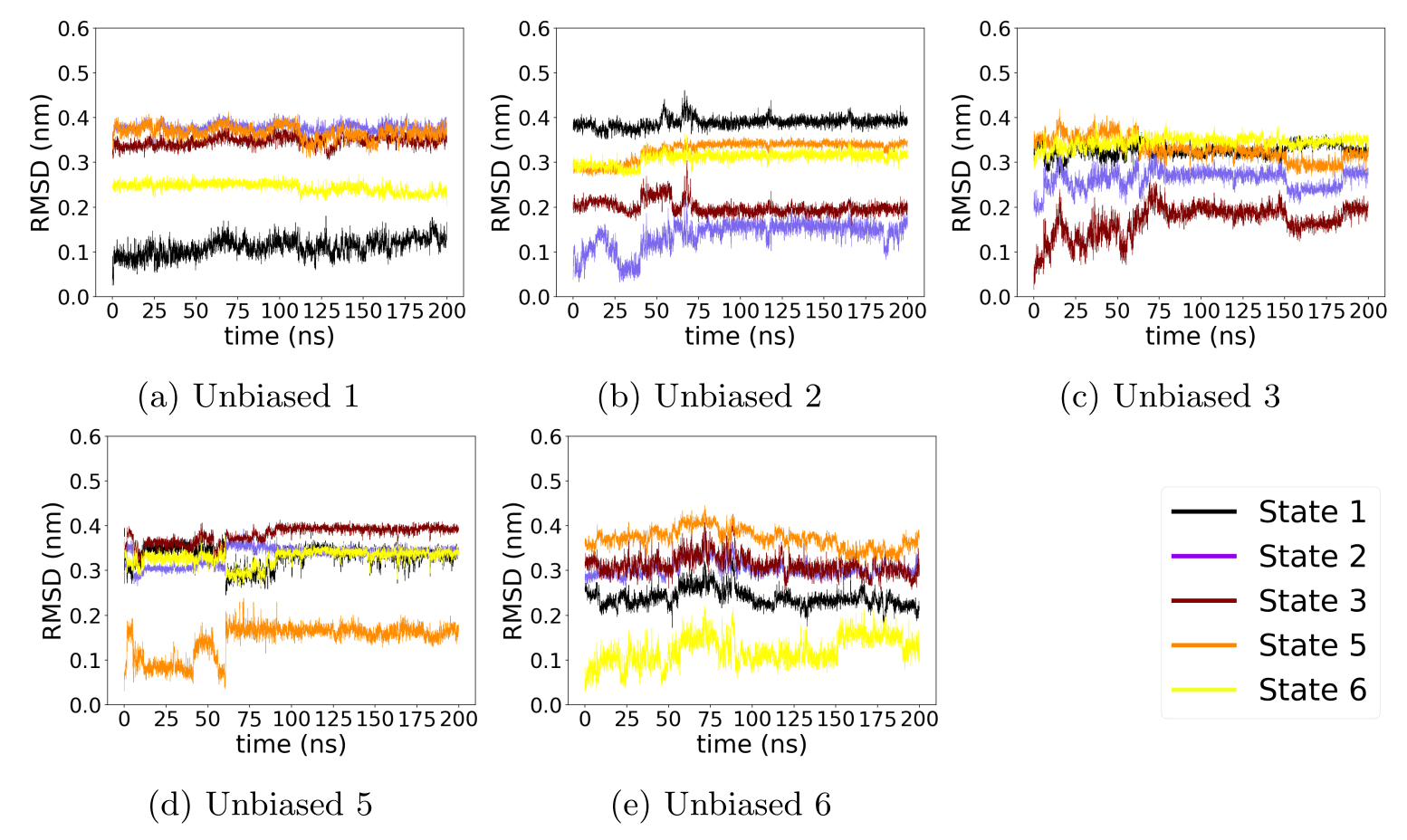}
\caption{RMSD of the $C_\alpha$ carbons of the L2 loop over a 200-ns unbiased MD trajectory for each of the five states.}
\end{figure}

\pagebreak
\newpage

\begin{figure}[h!]
\centering
\includegraphics[width=0.95\textwidth]{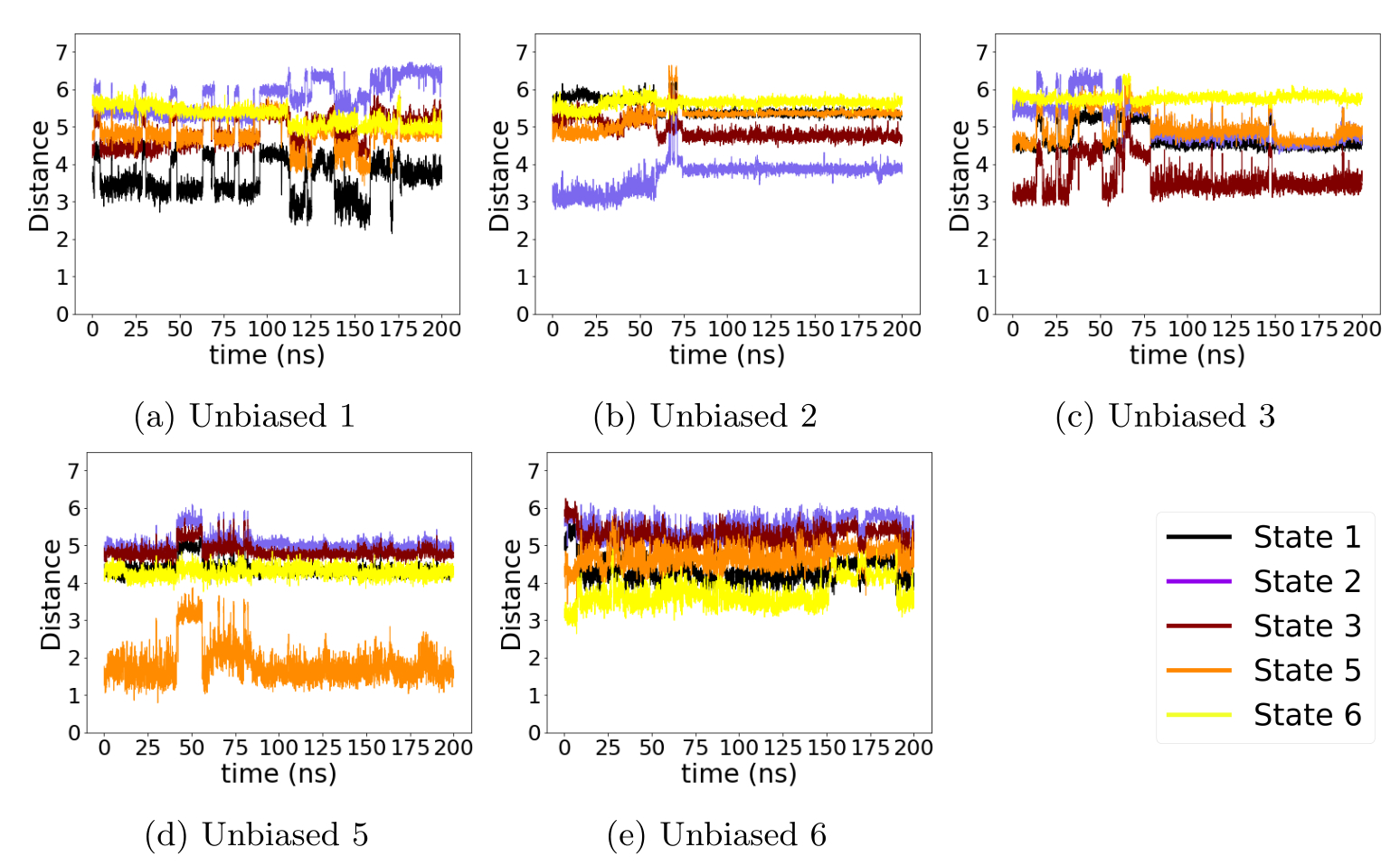}
\caption{Cluster centroid distances computed over the 200-ns unbiased MD trajectories started from the the five predefined Hsp90-NTD states.}
\end{figure}

\pagebreak
\newpage

\begin{figure}[h!]
\centering
\includegraphics[width=0.95\textwidth]{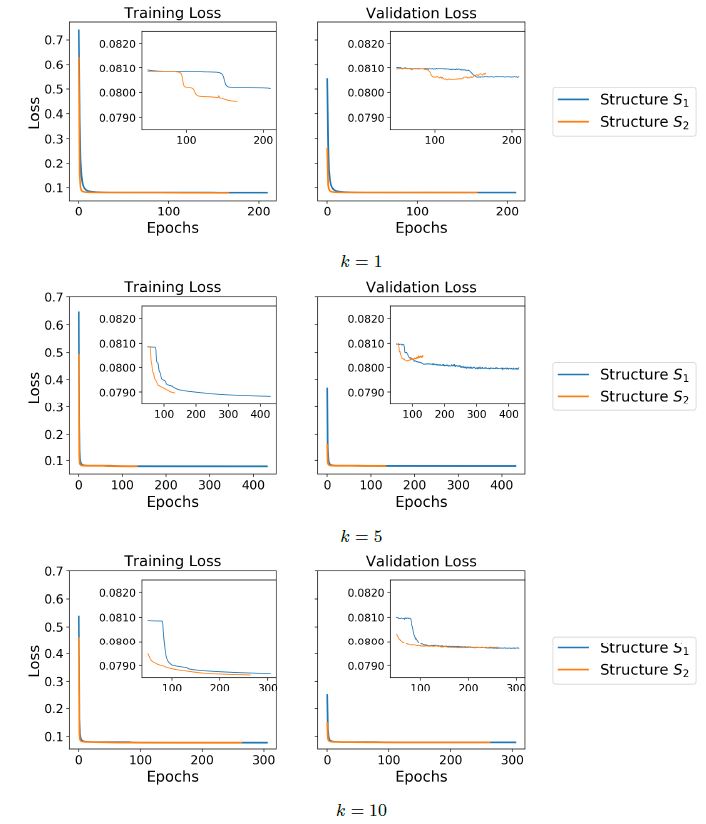}
\caption{Evolution of training and validation losses for two different autoencoder structures
and three different values of the bottleneck layer size $k$.}
\end{figure}

\pagebreak
\newpage

\begin{figure}[h!]
\centering
\includegraphics[width=0.7\textwidth]{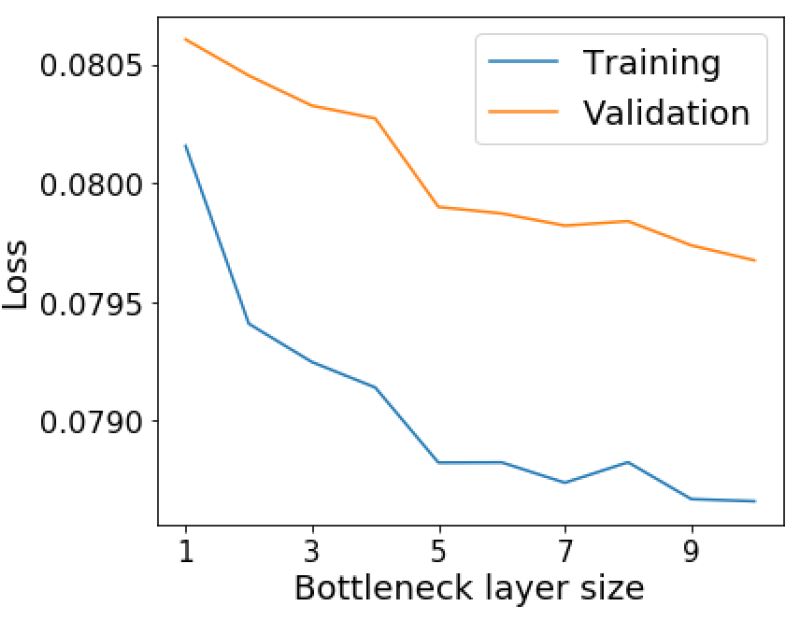}
\caption{Final training (blue) and validation (orange) loss obtained for each of the models corresponding to different $k$ bottleneck layer sizes.}
\end{figure}

\pagebreak
\newpage

\begin{figure}[h!]
\centering
\includegraphics[width=0.7\textwidth]{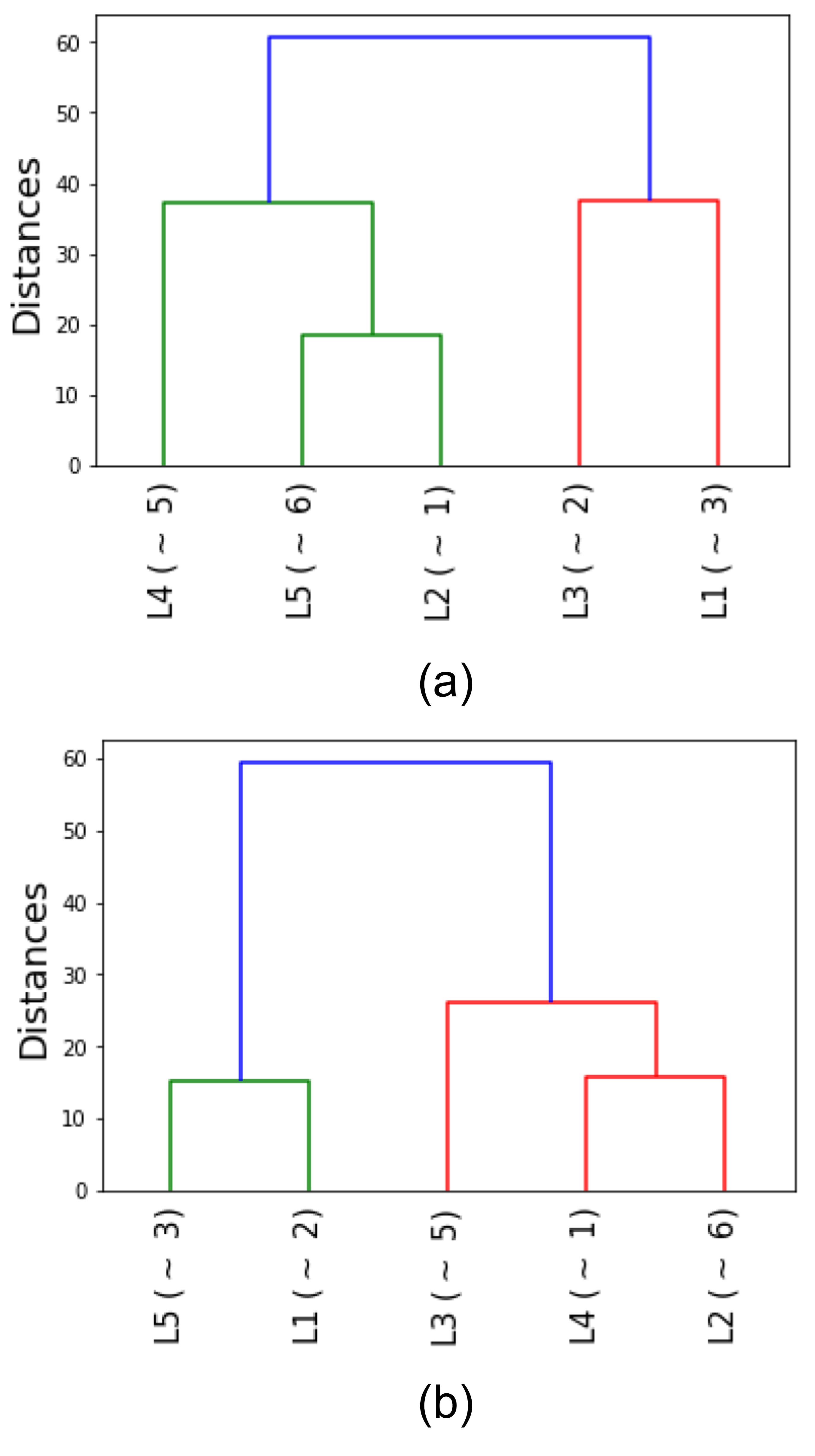}
\caption{Dendograms showing how the 5 clusters are iteratively merged into one. Each cluster is represented by its label number ($L_1$ to $L_5$) and the corresponding conformational state is indicated between brackets. The vertical axis corresponds to the distances between pairs of clusters. (a) 5-dimensional CV and (b) selected coordinate pair ($CV_3;CV_5$).}
\end{figure}

\pagebreak
\newpage

\begin{figure}[h!]
\centering
\includegraphics[width=0.6\textwidth]{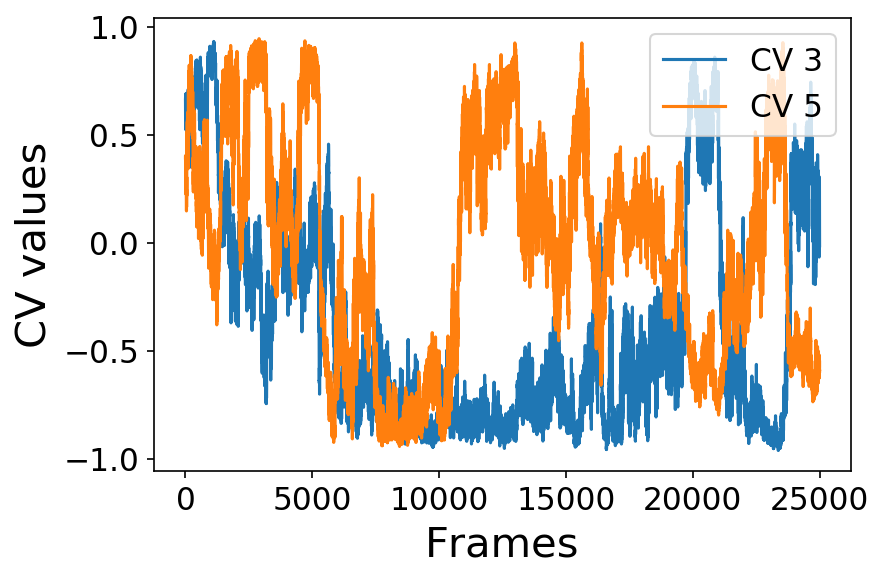}
\caption{Results of the 100 ns biased trajectory initiated from state 1 using ($CV_3;CV_5$) for biasing. Fluctuations of the two coordinates composing the biasing CV as a function of time.}
\end{figure}

\pagebreak
\newpage

\begin{figure}[h!]
\centering
\includegraphics[width=0.95\textwidth]{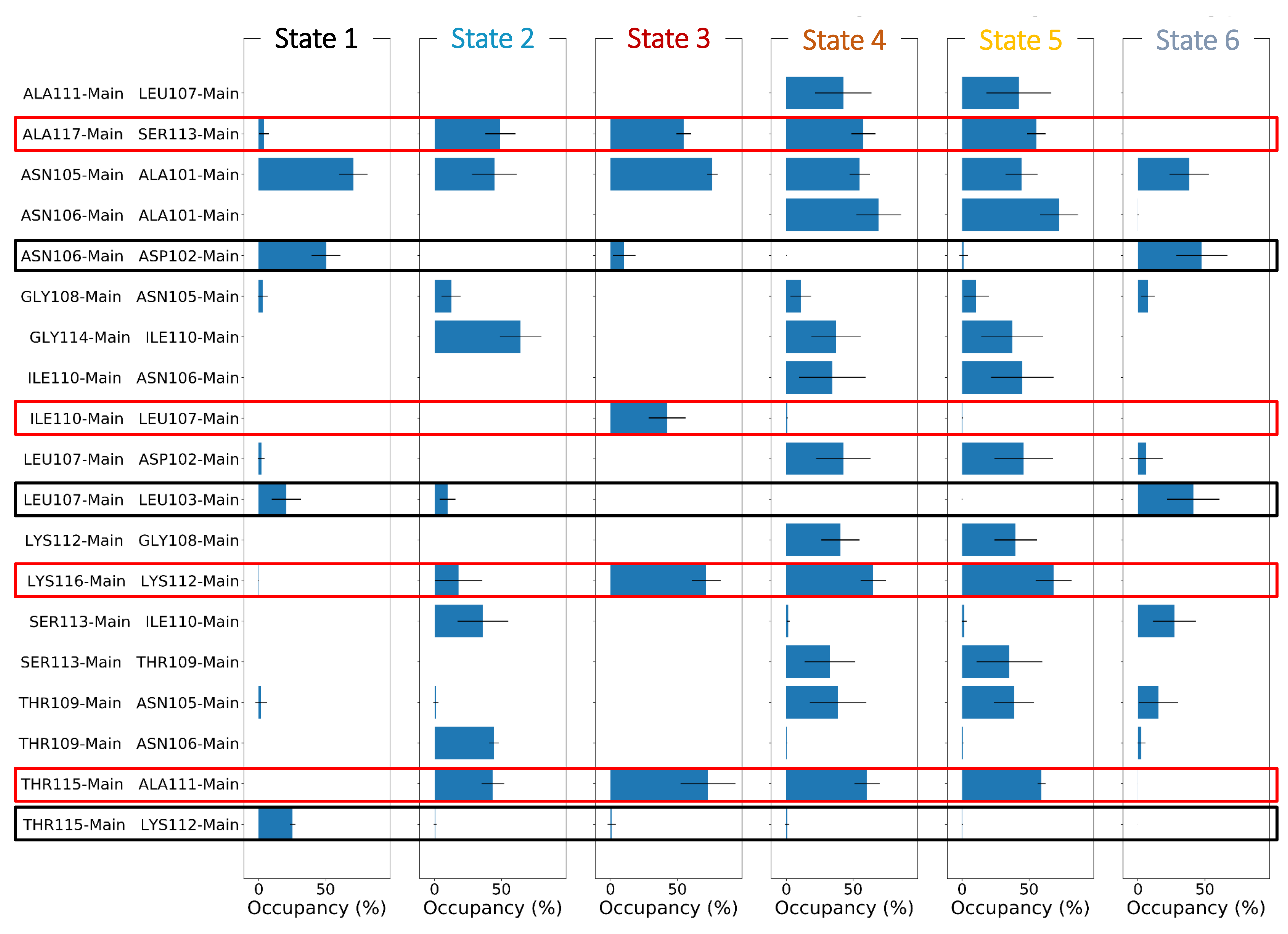}
\caption{Hydrogen bonds occurence percentages for all states using the initial unbiased MD dataset. The results for states 1 and 3 and highlighted in black and red, respectively}
\end{figure}

\pagebreak
\newpage

\begin{figure}[h!]
\centering
\includegraphics[width=0.95\textwidth]{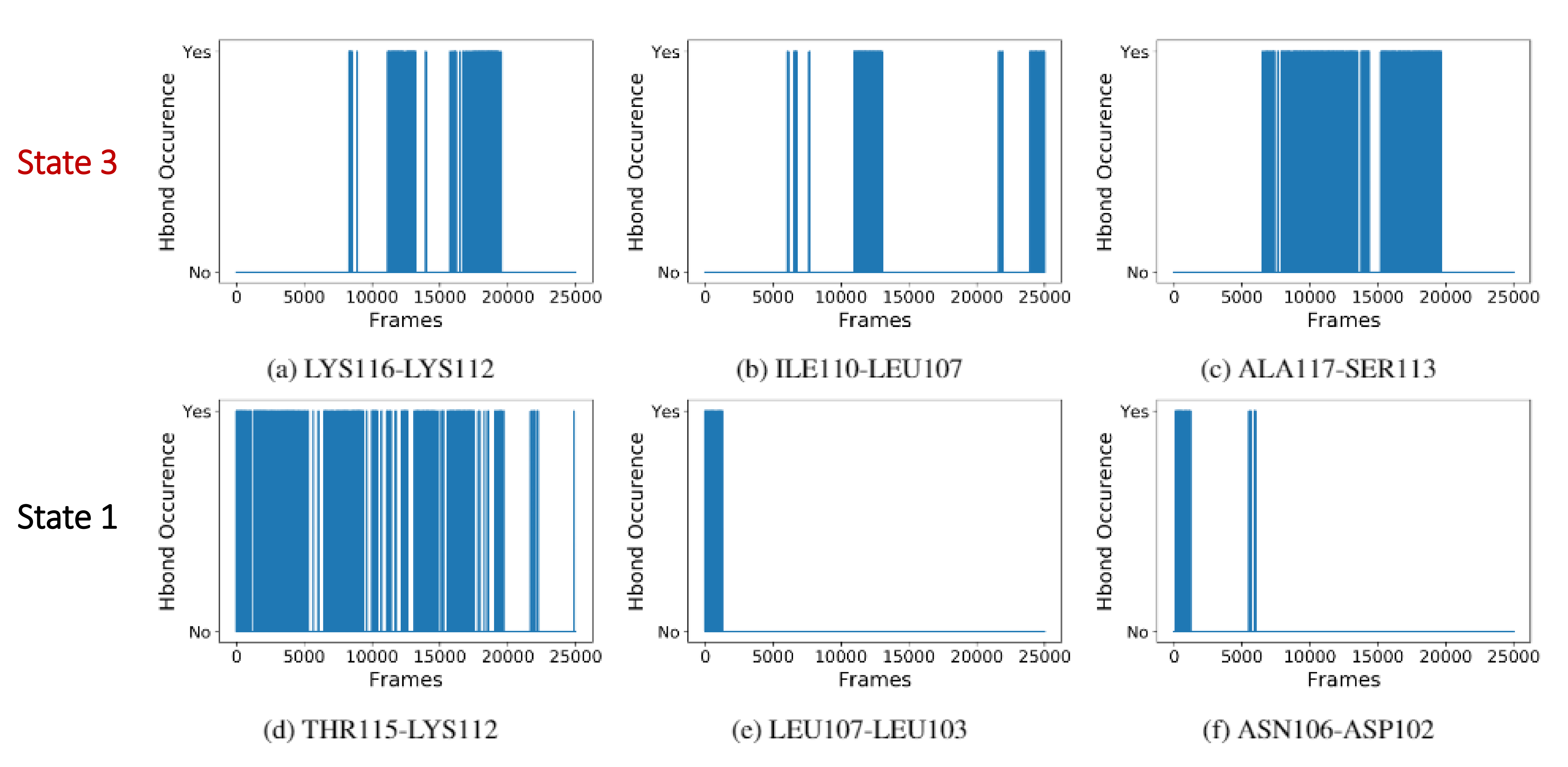}
\caption{Occurrences of the hydrogen bonds identified as high frequency and discriminant for State 1 and State 2. Results from the 100-ns biased MD simulation starting with a state 1 conformation.}
\end{figure}

\pagebreak
\newpage

\begin{figure}[h!]
\centering
\includegraphics[width=0.95\textwidth]{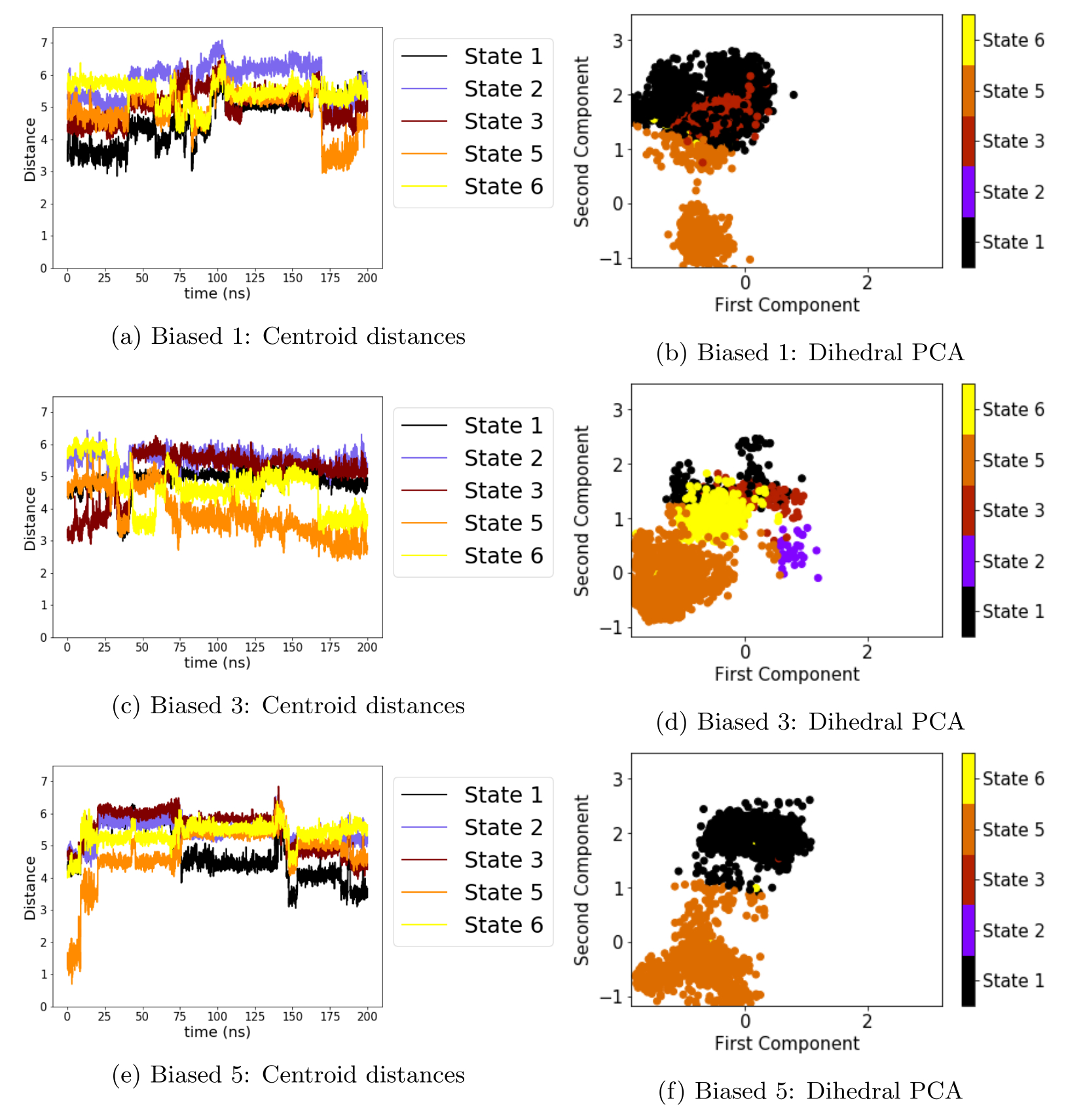}
\caption{Dihedral angle analysis of three additional 200-ns biased simulations starting from States 1, 3, and 5 and using the 2D-CV ($CV_3$;$CV_5$). Left: Distances to the centroids of the dihedral angle clustering. Right: Dihedral PCA projections.}
\end{figure}